\documentclass[11pt,a4paper,final]{revtex4}

\usepackage{sidecap}

\usepackage{ulem}
\usepackage{epsfig}
\usepackage{amsmath,amssymb,amsthm,mathtools}
\usepackage{graphicx}
\usepackage{bm}
\usepackage{color,soul}
\usepackage{latexsym}   
\usepackage{amsfonts} 	

\DeclareMathAlphabet{\mathpzc}{OT1}{pzc}{m}{it}

\begin{document}

\renewcommand{\textfraction}{0.00}


\newcommand{\pT}{p_\perp}
\newcommand{\vAi}{{\cal A}_{i_1\cdots i_n}}
\newcommand{\vAim}{{\cal A}_{i_1\cdots i_{n-1}}}
\newcommand{\vAbi}{\bar{\cal A}^{i_1\cdots i_n}}
\newcommand{\vAbim}{\bar{\cal A}^{i_1\cdots i_{n-1}}}
\newcommand{\htS}{\hat{S}}
\newcommand{\htR}{\hat{R}}
\newcommand{\htB}{\hat{B}}
\newcommand{\htD}{\hat{D}}
\newcommand{\htV}{\hat{V}}
\newcommand{\cT}{{\cal T}}
\newcommand{\cM}{{\cal M}}
\newcommand{\cMs}{{\cal M}^*}
\newcommand{\vk}{\vec{\mathbf{k}}}
\newcommand{\bk}{\bm{k}}
\newcommand{\kt}{\bm{k}_\perp}
\newcommand{\kp}{k_\perp}
\newcommand{\km}{k_\mathrm{max}}
\newcommand{\vl}{\vec{\mathbf{l}}}
\newcommand{\bl}{\bm{l}}
\newcommand{\bK}{\bm{K}}
\newcommand{\qm}{q_\mathrm{max}}
\newcommand{\vp}{\vec{\mathbf{p}}}
\newcommand{\bp}{\bm{p}}
\newcommand{\vq}{\vec{\mathbf{q}}}
\newcommand{\bq}{\textbf{q}}
\newcommand{\qt}{\bm{q}_\perp}
\newcommand{\qp}{q_\perp}
\newcommand{\bQ}{\bm{Q}}
\newcommand{\vx}{\vec{\mathbf{x}}}
\newcommand{\bx}{\bm{x}}
\newcommand{\tr}{{{\rm Tr\,}}}
\newcommand{\sNN}{s_{\mathrm{NN}}}
\newcommand{\bc}{\textcolor{blue}}

\newcommand{\beq}{\begin{equation}}
\newcommand{\eeq}{\end{equation}}
\newcommand{\eea}{\end{eqnarray}}
\newcommand{\beqar}{\begin{eqnarray}}
\newcommand{\eeqar}{\end{eqnarray}}

\newcommand{\half}{{\textstyle\frac{1}{2}}}
\newcommand{\ben}{\begin{enumerate}}
\newcommand{\een}{\end{enumerate}}
\newcommand{\bit}{\begin{itemize}}
\newcommand{\eit}{\end{itemize}}
\newcommand{\ec}{\end{center}}
\newcommand{\bra}[1]{\langle {#1}|}
\newcommand{\ket}[1]{|{#1}\rangle}
\newcommand{\norm}[2]{\langle{#1}|{#2}\rangle}
\newcommand{\brac}[3]{\langle{#1}|{#2}|{#3}\rangle}
\newcommand{\hilb}{{\cal H}}
\newcommand{\pleft}{\stackrel{\leftarrow}{\partial}}
\newcommand{\pright}{\stackrel{\rightarrow}{\partial}}

\newcommand{\meqbox}[2]{\eqmakebox[#1]{$\displaystyle#2$}}
\newcommand{\bb}[1]{ {\mathbf #1 }}

\newcommand{\kT}{\textbf{k}}
\newcommand{\qT}{\textbf{q}_{1}}
\newcommand{\qTd}{\textbf{q}_{2}}
\newcommand{\qTt}{\textbf{q}_{3}}
\newcommand{\qTc}{\textbf{q}_{4}}
\newcommand{\kTs}{k_{\perp}}
\newcommand{\qTs}{q_{1\perp}}
\newcommand{\qTds}{q_{2\perp}}
\newcommand{\qTts}{q_{3\perp}}
\newcommand{\qTcs}{q_{4\perp}}


\title{Importance of higher orders in opacity in QGP tomography}

\author{Stefan Stojku}
\affiliation{Institute of Physics Belgrade, University of Belgrade, Serbia}

\author{Bojana Ilic}
\affiliation{Institute of Physics Belgrade, University of Belgrade, Serbia}

\author{Igor Salom}
\affiliation{Institute of Physics Belgrade, University of Belgrade, Serbia}

\author{Magdalena Djordjevic\footnote{E-mail: magda@ipb.ac.rs}}
\affiliation{Institute of Physics Belgrade, University of Belgrade, Serbia}

\begin{abstract}
We consider the problem of including a finite number of scattering centers in dynamical energy loss and classical DGLV formalism. Previously, either one or an infinite number of scattering centers were considered in energy loss models, while efforts to relax such approximations require a more conclusive and complete treatment. In reality, however, the number of scattering centers is generally estimated to be 4-5 at RHIC and the LHC, making the above approximations (a priori) inadequate and this theoretical problem significant for QGP tomography.

We derived explicit analytical expressions for dynamical energy loss and DGLV up to the $4^{th}$ order in opacity, resulting in complex - highly oscillatory - mathematical expressions. These expressions were then implemented into an appropriately generalized DREENA framework to calculate the effects of higher orders in opacity on a wide range of high-$p_\perp$ light and heavy flavor predictions. Results of extensive numerical analysis and interpretations of nonintuitive results are presented. We find that, for both RHIC and the LHC, higher-order effects on high-$p_\perp$ observables are small, and the approximation of a single scattering center is adequate for dynamical energy loss and DGLV formalisms.
\end{abstract}

\maketitle

\section{Introduction}

Quark-Gluon Plasma (QGP)~\cite{QGP1,QGP2,QGP3,QGP4} is a new form of matter consisting of quarks, antiquarks, and gluons that are no
longer confined. It can be created in landmark experiments - RHIC and the LHC (so-called Little Bangs), where heavy ions collide at ultra-relativistic energies~\cite{QGP2,QGP3}. 
Hard probes are one of the main tools for understanding and characterizing the QGP properties~\cite{QGP2}, where hard processes dominate interactions of these probes with QGP constituents. These interactions are dominantly described by energy loss, where radiative is one of the most important mechanisms at high transverse momentum ($p_\perp$). The radiative energy loss can be analytically computed through pQCD approaches, typically under the assumption of the optically thick or optically thin medium (e.g., BDMPS-Z~\cite{BDMPS,Z}, ASW~\cite{ASW}, (D)GLV~\cite{GLV,DG}, HT and HT-M~\cite{HT,HTM}, AMY~\cite{AMY}, dynamical energy loss~\cite{MD_PRC,DH_PRL} and different applications/extensions of these methods) and tested against the experimental data.

Optically thick medium corresponds to the approximation of a jet experiencing infinite scatterings with medium constituents. While such an approximation would be adequate for QGP created in the early universe (Big Bang), Little Bangs are characterized by short, finite-size droplets of QCD matter. Another widely used approximation is an optically thin medium, assuming one scattering center. However, the medium created in Little Bangs is typically several fm in size (with mean free path $\lambda \approx 1$~fm), so considering several scattering centers in energy loss calculations is needed. Thus, it is evident that both approaches represent two extreme limits to the realistic situations considered in RHIC and LHC experiments, and relaxing these approximations to the case of a finite number of scattering centers is necessary. Thus, relaxing such approximation is a highly nontrivial problem, first addressed in \cite{GLV}, with recently renewed interest~\cite{Andres,Andres2,Mehtar-Tani1,Mehtar-Tani2,Sievert1,Sievert2,Wicks}. Some of these approaches are analytically quite advanced, e.g., providing full expressions for a gluon radiation spectrum (or splitting functions) with relaxed soft-gluon approximation in DGLV formalism~\cite{Sievert1,Sievert2} or derivation of gluon emission spectrum with full resummation of multiple scatterings within the BDMPS-Z framework~\cite{Andres,Mehtar-Tani1,Mehtar-Tani2}. However, in our view, this issue requires a more conclusive and complete treatment. Namely, the importance of including higher orders in opacity effects on experimental observables is still not addressed. In relaxing this approximation, it is not only needed to estimate these effects on, e.g., the energy loss and gluon radiation spectrum, but also to implement these corrections in the numerical frameworks needed to generate predictions for high-$p_{\perp}$ observables measured at RHIC and the LHC experiments. Furthermore, most of these studies were done in massless quarks and gluons limit and/or use the approximation of an uncorrelated medium (i.e., where the spacings between collisions are considered to be mutually independent, see~\cite{Wicks} for more details). Since we, {\it a priori}, do not know the magnitude of the effects of the inclusion of multiple scattering centers, nor how the mentioned approximations can influence this magnitude, we find it questionable to discuss higher order corrections while ignoring the effects which might potentially overshadow or alter the final effects. For example, due to a finite temperature medium, light quarks and gluons gain mass in QGP, which can significantly numerically modify the importance of these effects on experimental observables.

In this study, we start from our dynamical energy loss formalism~\cite{MD_PRC,DH_PRL}, computed under the approximation of an optically thin QCD medium, i.e., one scattering center. We use general expressions from~\cite{Wicks} to relax this approximation to the case of finite number of scattering centers, where explicit analytical expressions up to the $4^{th}$ order in opacity (scattering centers) are presented. These expressions are implemented in our (appropriately modified) DREENA-C~\cite{DREENA_C} framework (which assumes a constant temperature medium), enabling us to more straightforwardly estimate the effects of higher orders in opacity on high-$p_\perp$ $R_{AA}$ and $v_2$ observables. Based on these results, we also provide estimates for the fully evolving medium, while a rigorous study in this direction is left for future work.

While the initial expressions taken from~\cite{Wicks} were, strictly speaking, derived in the approximation of static scattering centers, we apply them here in the context of a dynamic QCD medium. Namely, by careful calculation, we have shown in~\cite{DH_PRL} that -- at least in the first order in opacity -- the generalization from the static to dynamic medium eventually amounts to a mere appropriate replacement of the mean free path and effective potential in the final expressions. Following general arguments given in~\cite{GLV} and the expectations expressed in~\cite{Wicks}, we assume that the same prescription for progressing from static to dynamic medium remains valid in higher orders of opacity.

The outline of the manuscript is as follows: Sections II and III present the outline of theoretical and numerical frameworks used in this study, with more detailed analytical results presented in the Appendices. In the Results section, we will numerically analyze the effects of higher orders in opacity on the gluon radiation spectrum and high-$p_\perp$ $R_{AA}$ and $v_2$ predictions. Intuitive explanations behind obtained results will be presented. This section will also analyze a special case of static QCD medium (extension of (D)GLV~\cite{GLV,DG} to the finite number of scattering centers). The main results will be summarized in the last section.

\section{Theoretical framework}

In this study, we use our dynamical radiative energy loss~\cite{MD_PRC,DH_PRL} formalism, which has the following features: {\it i)} QCD medium of {\it finite} size $(L)$ and temperature ($T$), which consists of dynamical (i.e., moving) partons, in a distinction to models with widely used static approximation and/or vacuum-like propagators~\cite{BDMPS,ASW,GLV,HT}. {\it ii)} Calculations based on generalized Hard-Thermal-Loop approach~\cite{Kapusta,Le_Bellac}, with naturally regulated infrared
divergences~\cite{MD_PRC,DH_PRL,DG_TM}. {\it iii)} Generalization towards running coupling~\cite{MD_PLB} and finite magnetic mass~\cite{MD_MagnMass}.

However, as noted in the Introduction, this radiative energy loss is developed up to the first order in opacity. Thus, to improve the applicability of this formalism for QGP tomography, it is necessary to relax this approximation. To generalize the dynamical energy loss to finite number in scattering centers, we start from a closed-form expression - Eq. (46) from~\cite{Wicks} and Eq. (20) from ~\cite{DG} - derived for static QCD medium (i.e., (D)GLV case~\cite{GLV,DG}) but applicable for a generalized form of effective potential and mean free path $\lambda$~\cite{Wicks}.
\beqar
\label{eqn:allorders}
 x \frac{dN^{(n)}}{dx\, d^2 {\bf k}} &=&
\int_0^L dz_1 \cdots \int_{z_{n-1}}^L dz_n \int \prod_{i=1}^n \left( d^2{\bf q}_{i}\, \frac{ v^2({\bf q}_{i})- \delta^2({\bf q}_{i}) }{\lambda(z)} \right) \nonumber \\
& \times& \frac{C_R \alpha_s(Q^2_k)}{\pi^2} \Bigl( -2\,{\bf C}_{(1 \cdots n)} \cdot {\bf B}_{n}
\left[ \cos \sum_{k=2}^n \omega_{(k\cdots n)} \Delta z_k
-   \cos \sum_{k=1}^n \omega_{(k\cdots n)} \Delta z_k
\right]\; \Bigr),
\eeqar
where $|v_i({\bf q}_{i}) |^2$ is defined as the normalized distribution of momentum transfers from the $i^{{\rm th}}$ scattering center (i.e., "effective potential"), $\lambda(i)$ is the mean free path of the emitted gluon, $C_R$ is the color Casimir of the jet. Note that, for consistency with our previous work, we denote transverse 2D vectors as bold ${\bf p}$.

The running coupling is defined as in~\cite{MD_PLB}:
\beqar
\alpha_s(Q^2) = \frac{4\pi}{(11-2/3n_f) \ln(Q^2/\Lambda_{QCD})},
\eeqar
where $Q^{2}_k= \frac{\textbf{k}^2+M^2x^2+m_{g}^2}{x}$, appearing in Eq.~(\ref{eqn:allorders}) above is the off-shellness of the jet before gluon radiation~\cite{MD_PLB}.

$\omega_{(m \ldots n)}$ is the inverse of the formation time or the (longitudinal) momentum
\begin{equation}
 \omega_{(m \ldots n)} = \frac{\chi^2 + (\bb{k}-\bb{q}_{m} - \ldots \bb{q}_{n})^2}{2 x E},
\end{equation}
where $n$ is the final scatter, while $m$ varies from the first up to the final scatter. $\chi^2\equiv M^2 x^2 + m_g^2$, where $x$ is the longitudinal momentum fraction of the quark jet carried away by the emitted gluon, $M$ is the mass of the quark, $m_g=\mu_E/\sqrt 2$ is the effective mass for gluons with hard momenta~\cite{DG_TM}, and $\mu_E$ is the Debye mass (i.e., electric screening).

`Cascade' terms represent the shifting of the momentum of the radiated gluon due to momentum kicks from the medium:
\begin{equation}
 \bb{C}_{(i_1 i_2 \ldots i_m)} = \frac{(\bb{k}-\bb{q}_{i_1} - \bb{q}_{i_2} - \ldots - \bb{q}_{i_m} )}{ \chi^2 + (\bb{k}-\bb{q}_{i_1} - \bb{q}_{i_2} - \ldots - \bb{q}_{i_m} )^2 }.
\end{equation}

A special case of $\bb{C}$ without any momentum shifts is defined as the `Hard' term:
\begin{equation}
 \bb{H} = \frac{\bb{k}}{\chi^2 + \bb{k}^2}, \,\, \mathrm{and } \,\, \bb{B}_i = \bb{H} - \bb{C}_i.
\end{equation}


In~\cite{MD_PRC,DH_PRL,MD_MagnMass}, we showed that, despite much more involved analytical calculations, at first order in opacity the radiative energy loss in a dynamical medium has the same form as in the static medium, except for two straightforward substitutions in mean free path and effective potential:
\beqar
\label{lambda_stat}
\lambda_\mathrm{stat} \rightarrow \lambda_\mathrm{dyn},
\eeqar
where $\lambda_\mathrm{stat}^{-1}= 6 \frac{1.202}{\pi^2} \frac{1+n_f/4}{1+n_f/6} \lambda_\mathrm{dyn}^{-1}$, while the 'dynamical mean free path' is given by $\lambda_{dyn}^{-1}=3\alpha_s(Q^2_v)T$~\cite{MD_PRC,DH_PRL}, with  $Q^2_v=ET$~\cite{MD_PLB}. Running coupling $\alpha_s(Q_v^2)$ corresponds to the interaction between the jet and the virtual (exchanged) gluon, while $E$ is the jet's energy.

\beqar
\label{diff_cross}
 \left[\frac{\mu_E^2}{\pi(\bb{q}^2{+}\mu_E^2)^2}\right]_\mathrm{stat}
   \rightarrow
 \left[ \frac{\mu_E^2-\mu_M^2}{\pi(\bb{q}^2{+}\mu_E^2)(\bb{q}^2{+}\mu_M^2) } \right]_\mathrm{dyn},
\eeqar
where $\mu_M$ is magnetic screening. Thus, we assume that the Eq.~(\ref{eqn:allorders}) can also be used in our case, with the above modification of effective potential and mean free path. In the Appendices A and B, we use this general expression to derive an explicit expression for the gluon radiation spectrum for $1^{\mathrm{st}}$, $2^{\mathrm{nd}}$, $3^{\mathrm{rd}}$ and $4^{\mathrm{th}}$ order in opacity ($\frac{dN_g^{(1)}}{dx}$, $\frac{dN_g^{(2)}}{dx}$, $\frac{dN_g^{(3)}}{dx}$, $\frac{dN_g^{(4)}}{dx}$, respectively).

\section{Numerical framework}

To generate the results presented in this work, we used our (appropriately generalized, see below) DREENA-C framework. For completeness, we here give a brief outline of this framework, while a detailed description is presented in~\cite{DREENA_C}. The quenched spectra of light and heavy quarks are calculated according to the generic pQCD convolution given by:
\begin{equation}
        \frac{E_fd^3\sigma}{dp_f^3}=\frac{E_id^3\sigma(Q)}{dp_i^3}\otimes P(E_i\rightarrow E_f)\otimes D(Q\rightarrow H_Q).
\end{equation} \par

Here, indices \textit{i} and \textit{f} stand for 'initial' and 'final', respectively, while $Q$ denotes initial high-energy parton (light quarks, heavy quarks, or gluons). $E_id^3\sigma(Q)/dp_i^3$ is the initial momentum spectrum for the given parton, which is calculated according to~\cite{Vitev0912}, $P(E_i \rightarrow E_f)$ represents the energy loss probability for the given particle which was calculated within the dynamical energy loss formalism~\cite{MD_PRC,DH_PRL}, which includes multi-gluon~\cite{GLV_suppress} and path-length fluctuations~\cite{WHDG,DREENA_C}. $D(Q \rightarrow H_Q)$ represents the fragmentation function of light and heavy partons into hadrons, where for light hadrons, D and B mesons, we use DSS~\cite{DSS}, BCFY~\cite{BCFY}, KLP~\cite{KLP} fragmentation functions, respectively.
The geometry is averaged over by using path-length distributions, i.e., probability distributions of the path lengths of hard partons in Pb+Pb collisions, in the same way as in the original DREENA-C framework~\cite{DREENA_C}. They are used as weight functions when integrating over the path-length in our numerical procedure.

We use the following parameters in the numerical procedure: $\Lambda_{QCD}=0.2$ GeV and $n_f = 3$. Temperature-dependent Debye chromoelectric mass $\mu_E (T)$ has been extracted from~\cite{Peshier}. For the mass of light quarks, we take the thermal mass $M \approx \mu_E / \sqrt{6}$, and for the gluon mass, we use $m_g = \mu_E / \sqrt{2}$ ~\cite{DG_TM}. The mass of the charm (bottom) quark is $M=1.2$ GeV ($M=4.75$ GeV). The magnetic and electric mass ratio is $0.4 < \mu_M / \mu_E < 0.6$~\cite{Maezawa,Nakamura}. All the results presented in this paper are generated for the Pb+Pb collision system at $\sqrt{s_{NN}}=5.02$ TeV.

As DREENA-C~\cite{DREENA_C} does not include suppression from multiple scattering centers in the medium, we now upgrade this framework to include the $2^{nd}$ and $3^{rd}$ order in opacity contributions. We integrate the expressions obtained from~(\ref{eqn:allorders}) analytically for $z_i$ (see Appendices A and B), and then numerically for momenta $\textbf{k}$ and $\textbf{q}_{i}$ using the quasi-Monte Carlo method to obtain $\frac{dN_g}{dx}$ up to $3^{rd}$ order in opacity. Also, to test the importance of multiple scattering centers on radiative energy loss, we exclude the collisional~\cite{MD_Coll} contributions from the DREENA-C framework and only generate predictions for radiative energy loss. Appendices A and B also include expressions for the $4^{th}$ order in opacity. We implemented $4^{th}$ order into DREENA-C, but as the resulting integrals are highly oscillatory, we could not reach convergence for this order using our available computational resources. Notably, this numerical complexity is significantly higher, estimated to be $\sim 2$ orders of magnitude larger than for the $3^{rd}$ order (e.g., for the $1^{st}$ order, we needed $\sim 25$ CPUh; for the $2^{nd}$ order $\sim 2500$ CPUh; for the $3^{rd}$ order $\sim 70000$ CPUh). Nevertheless, at specific points where we reached a convergence, we found the $4^{th}$ order contribution negligible, as expected from the results presented in the next section.

\section{Results}
\begin{figure*}
\centering
\epsfig{file=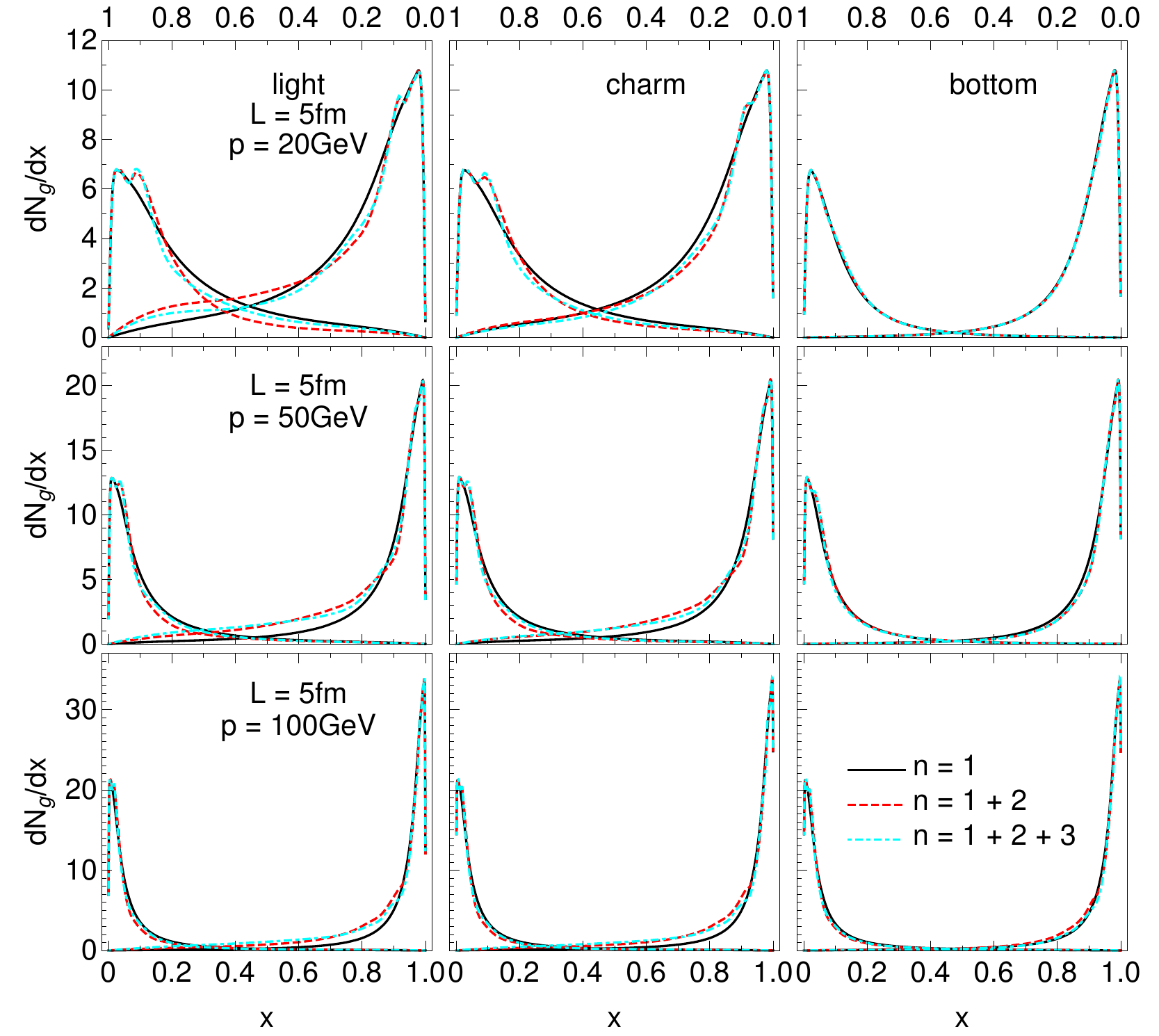,scale=0.4}
\vspace*{-0.2cm}
\caption{Gluon radiation spectrum $dN_g/dx$ as a function of $x$, for the typical medium length of $L=5fm$ and various jet momenta. Different columns correspond to light, charm, and bottom quarks. Solid black curves show the $1^{\mathrm{st}}$ order in opacity results, red dashed curves show the results up to the $2^{\mathrm{nd}}$ order, while cyan dot-dashed curves up to the $3^{\mathrm{rd}}$ order in opacity. Curves with the peaks on the left (right) side of each of the plots correspond to the $\mu_M / \mu_E = 0.6$ ($\mu_M / \mu_E = 0.4$) case, and the numerical values should be read off on the lower (upper) $x$-axis.}
\label{fig:dNdx}
\end{figure*}

In Fig.~\ref{fig:dNdx}, the effect of higher orders in opacity on $\frac{dN_g}{dx}$ as a function of $x$ is shown for typical medium length $L=5$fm. In each plot, we use double axes for clarity: the lower axis corresponds to magnetic to electric mass ratio $\mu_M/\mu_E$=0.6 (and the curves with the peak on the left side), while the upper axis corresponds to $\mu_M/\mu_E$=0.4 (and the curves with the peak on the right side) - note that, in each case, maximum is reached for low values of $x$. We see that the importance of higher orders of opacity decreases with the increase of jet energy and mass. They also decrease with decreasing the size of the medium, as shown in the Appendix C (equivalent figures for $L=3$fm (Fig.~\ref{fig:dNdxL3}, left) and $L=1$fm (Fig.~\ref{fig:dNdxL3}, right)). For bottom quarks, higher-order effects are negligible independently of the jet momentum. In contrast, these effects are moderate for charm and light quarks and can influence the jet observables, as discussed below.
Note that, due to color triviality, the results for light quarks show the (scaled) result for gluons, too. This holds up to the fact that, due to the indistinguishability of the radiated gluon from the gluon in the jet, the limits for subsequent integration of $dN_g/dx$ with respect to $x$ is performed from $x_{lower} = 0$ to $x_{upper} = 1/2$ (as opposed to $x_{upper}=1$ for light quarks).

\begin{figure*}
\centering
\epsfig{file=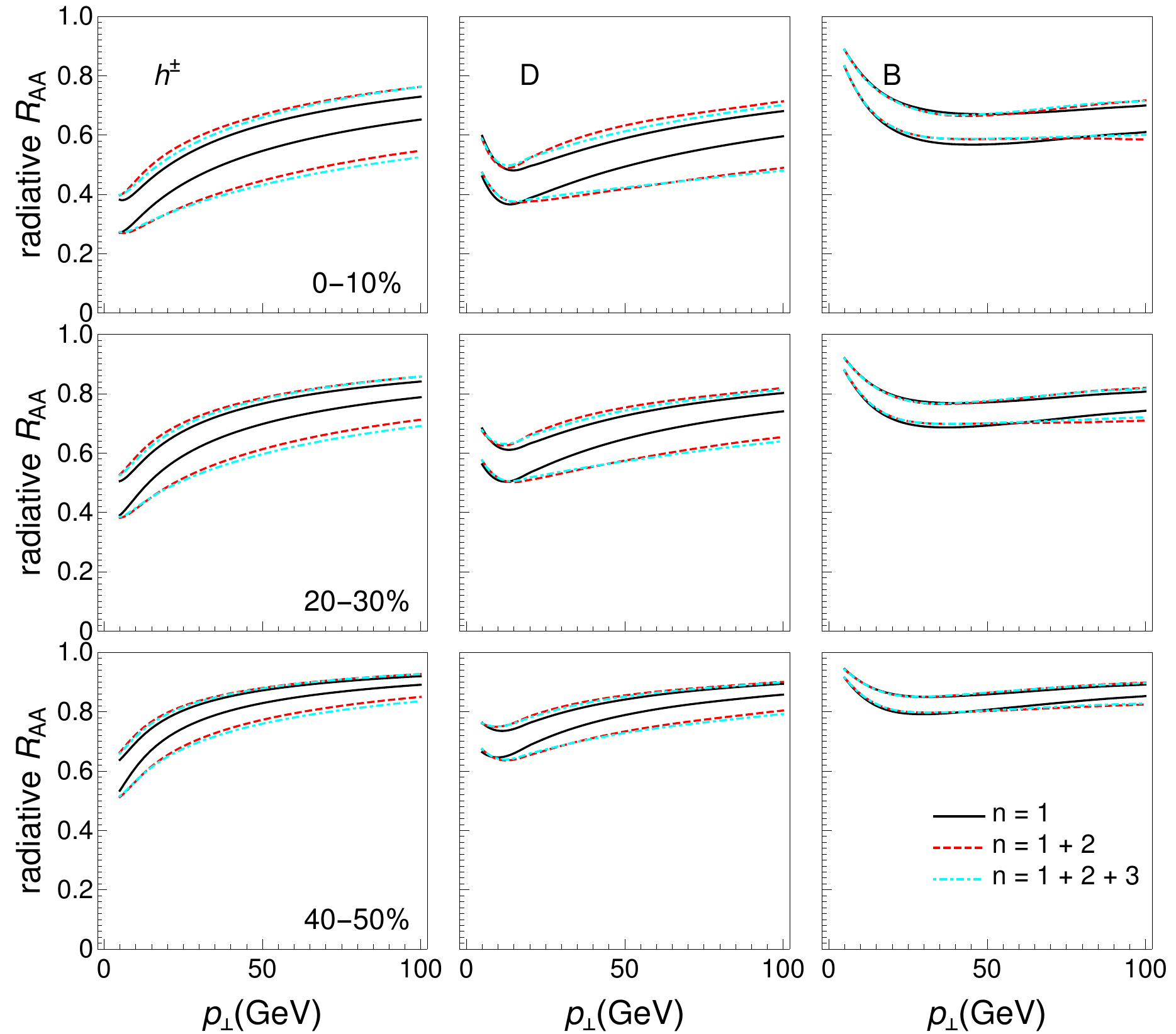,scale=0.4}
\vspace*{-0.2cm}
\caption{Radiative $R_{AA}$ results obtained within DREENA-C -- the effects of different orders in opacity. The results are generated for the Pb+Pb collision system at $\sqrt{s_{NN}}=5.02$ TeV, and all the other figures in the manuscript show the results for the same collision system and energy.  Different columns correspond to charged hadrons, D, and B mesons, while different rows show different centrality classes. Solid black curves show the $1^{\mathrm{st}}$ order in opacity results, red dashed curves show the results up to the $2^{\mathrm{nd}}$ order, while cyan dot-dashed curves up to the $3^{\mathrm{rd}}$ order in opacity. The upper (lower) boundary of each band corresponds to the $\mu_M / \mu_E = 0.6$ ($\mu_M / \mu_E = 0.4$) case.}
\label{fig:RAA}
\end{figure*}

In Fig.~\ref{fig:RAA}, we show the effect of higher orders in opacity on radiative $R_{AA}$ observable. Our computations have shown that the effect on $v_2$ is similar to the one on $R_{AA}$ (see Fig.~\ref{fig:v2} in Appendix D). Thus, to avoid redundancy, we further concentrate only on $R_{AA}$.

We first observe that the effect on $R_{AA}$ is smaller for more peripheral collisions. This is expected, as the medium is shorter on average, so including multiple scattering centers becomes less important.

Furthermore, we find that higher orders in opacity are negligible for B mesons, while these effects increase with decreasing mass, as expected from Fig.~\ref{fig:dNdx}. The reason behind this is the decrease in the gluon formation time with increasing jet mass. When the gluon formation time is short, the energy loss approaches the incoherent limit, where it was previously shown that the effects of higher orders in opacity are negligible~\cite{DG}. Thus, our results are consistent with the previous findings. On the other hand, for large gluon formation time (massless quark and gluon limit), the higher orders in opacity effects become significant, also in general agreement with the previous findings~\cite{Andres}. In finite temperature QGP (considered in this study), light quarks and gluons gain mass due to Debye screening, reducing the effects of higher orders in opacity on the energy loss, consistently with Fig.~\ref{fig:RAA}.

Unexpectedly, we also observe that, for different magnetic mass limiting cases, these effects on $R_{AA}$ are opposite in sign: for $\mu_M/\mu_E$=0.6, the inclusion of higher orders in opacity reduces energy loss (and, consequently, suppression). In contrast, for $\mu_M/\mu_E$=0.4, the effect is both opposite in sign and larger in magnitude. What is the reason behind these unexpected results?

To answer this question, we go back to the effective potential~\cite{MD_MagnMass} $v(\bq)$ in dynamical QCD medium, which can be written in the following form
\beqar
v(\bq) &=& v_L(\bq) - v_T(\bq) \, ,
\label{vLT}
\eeqar
where $v_L(\bq)$ is longitudinal (electric), and $v_T(\bq)$ is transverse (magnetic), contribution to the effective potential. The general expressions for the transverse and longitudinal contributions to
the effective potentials are
\beqar
v_{L}(\bq) &=& \frac{1}{\pi} \left(\frac{1}{(\bq^2+\mu_{pl}^2)} - \frac{1}{(\bq^2+\mu_{E}^2)} \right),
\quad v_{T}(\bq) = \frac{1}{\pi} \left(\frac{1}{(\bq^2+\mu_{pl}^2)} - \frac{1}{(\bq^2+\mu_{M}^2)} \right),
\label{vTLf}
\eeqar
where $\mu_{E}$, $\mu_{M}$ and $\mu_{pl}=\mu_{E}/\sqrt{3}$ are electric, magnetic and plasmon masses, respectively.
As seen from Eq.~(\ref{vLT}), this potential has two contributions - electric and magnetic, where the electric contribution is always positive due to $\mu_{pl} < \mu_{E}$. On the other hand, magnetic contribution depends non-trivially on the value of magnetic mass. That is, for $\mu_{M} > \mu_{pl}$, we see that magnetic contribution decreases the energy loss, while for $\mu_{M} < \mu_{pl}$ it increases the energy loss and consequently suppression, as shown in Fig.~\ref{fig:RAA}, which may intuitively explain the observed energy loss behavior.

\begin{figure*}
\centering
\epsfig{file=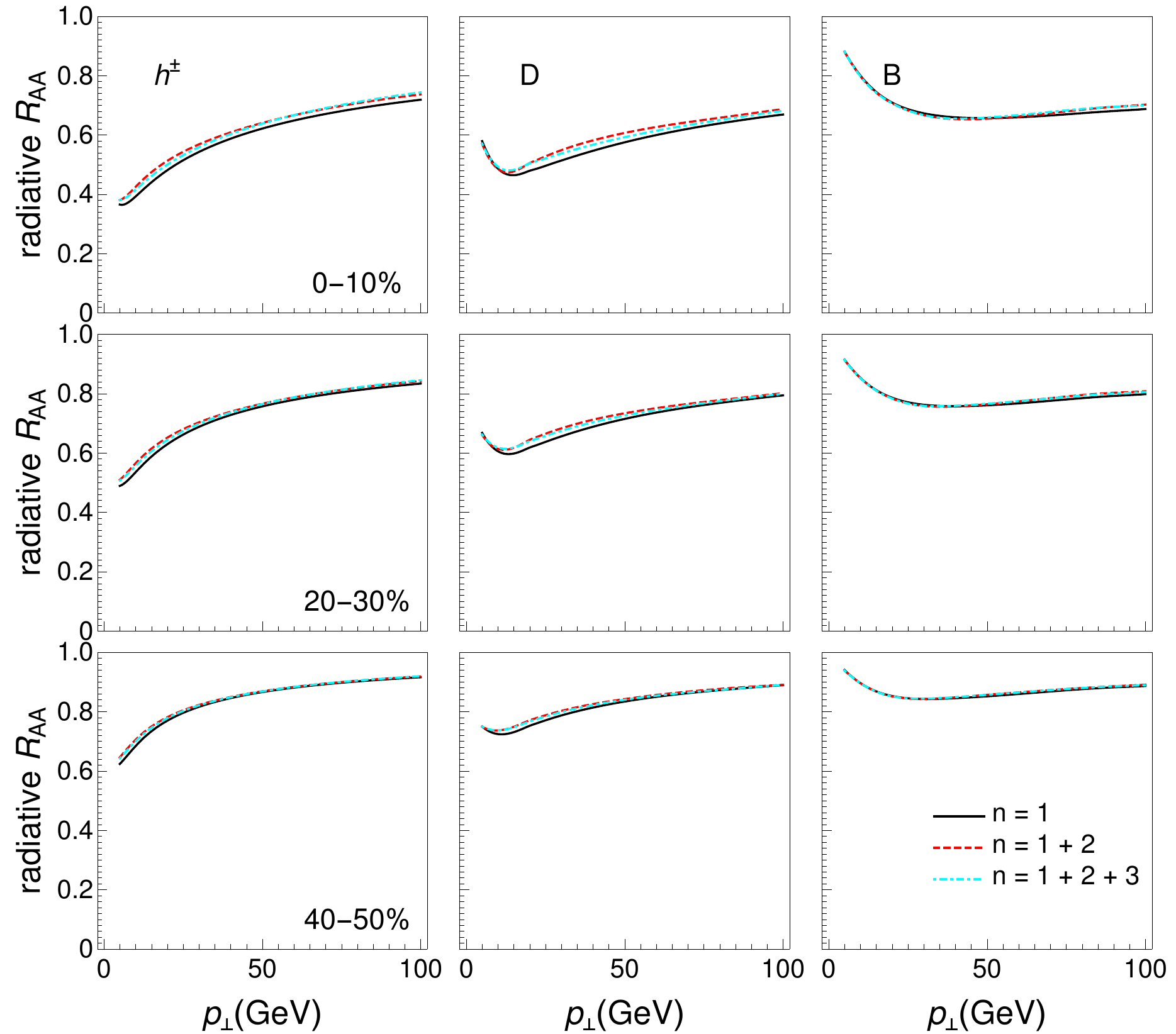,scale=0.4}
\vspace*{-0.2cm}
\caption{$R_{AA}$ results, obtained within DREENA-C when only electric contribution ($v_L(\bq)$) to radiative energy loss is considered. Different columns correspond to charged hadrons, D and B mesons, while different rows show different centrality classes. Solid black curves show the $1^{\mathrm{st}}$ order in opacity results, red dashed curves show the results up to the $2^{\mathrm{nd}}$ order, while cyan dot-dashed curves up to the $3^{\mathrm{rd}}$ order in opacity. }
\label{fig:RAA_el}
\end{figure*}

Furthermore, the Debye mass $\mu_E$ is well defined from lattice QCD, where the perturbative calculations are also consistent~\cite{Peshier}. Thus, the electric potential is well defined in dynamical energy loss, and we can separately test the effect of higher orders in opacity on this contribution (by replacing $v(\bq)$ by $v_L(\bq)$ in the DREENA framework). We surprisingly find it to be negligible, as shown in Fig.~\ref{fig:RAA_el}. Thus, higher orders in opacity essentially do not influence the electric contribution in a dynamical QCD medium, which is an interesting and intuitively unexpected result. That is, the higher orders mainly influence the magnetic contribution to energy loss (keeping the electric contribution unaffected), where the sign of the effect depends on the magnetic mass value. For example, as $\mu_M/\mu_E$=0.4 is notably smaller than $\mu_{pl}/\mu_E = 1/\sqrt{3}$, the higher orders in opacity are significant for this limit and increase the suppression, in agreement with Fig.~\ref{fig:RAA}. On the other hand, $\mu_M/\mu_E$=0.6 is close to (but slightly larger than) $\mu_{pl}/\mu_E$, so higher orders in opacity are small for this magnetic mass limit and reduce the suppression, also in agreement with Fig.~\ref{fig:RAA}. Additionally, note that the most recent 2+1 flavor lattice QCD results with physical quark masses further constrain the magnetic screening to $0.58<\mu_M/\mu_E<0.64$~\cite{Borsanyi:2015yka}. Thus, for this range of magnetic screening, we conclude that the effects of higher orders in opacity are small in a dynamical QCD medium and can be safely neglected.

\begin{figure*}
\centering
\epsfig{file=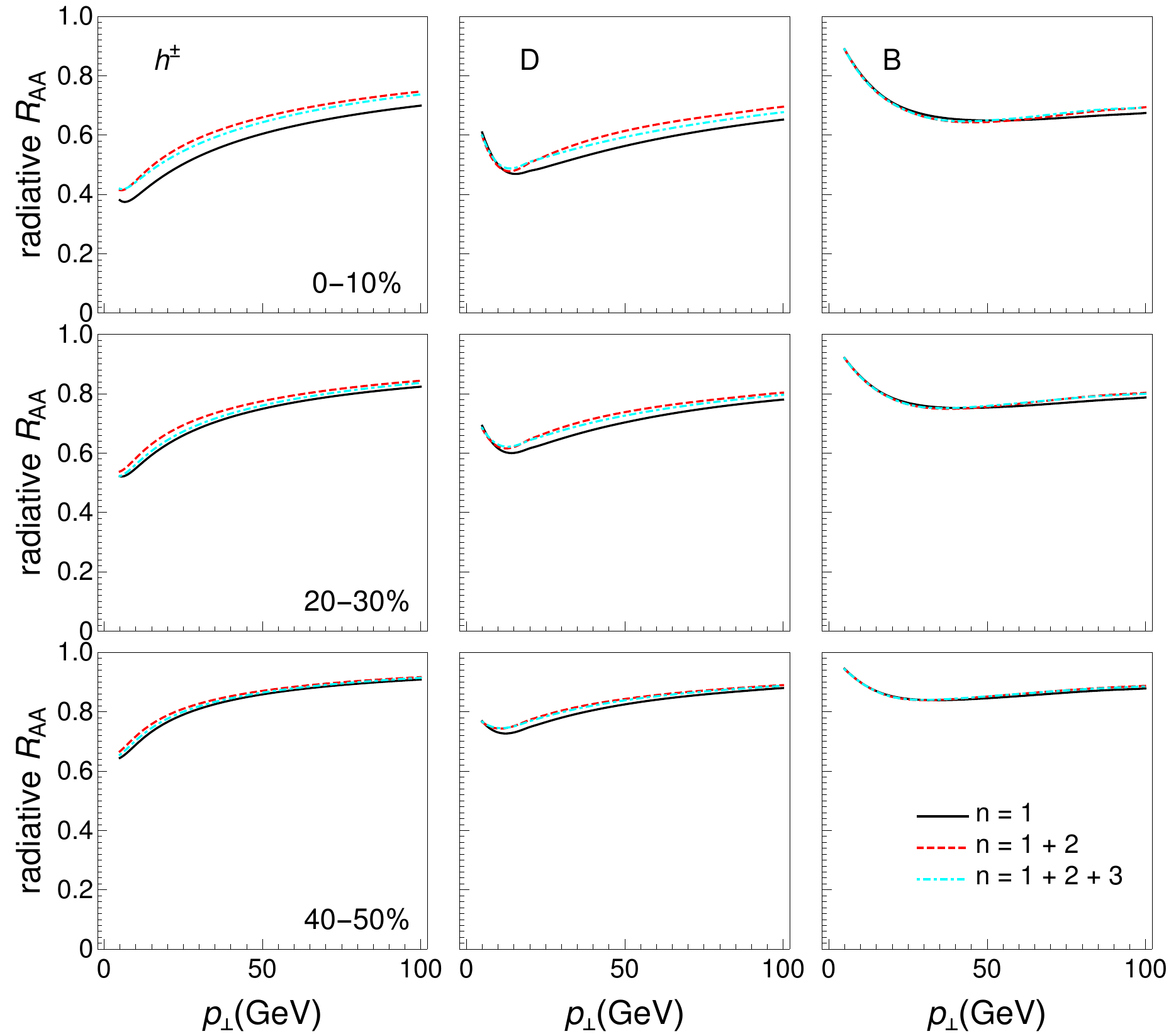,scale=0.4}
\vspace*{-0.2cm}
\caption{Radiative $R_{AA}$ results obtained within DREENA-C under the \textit{static} medium approximation. Different columns correspond to charged hadrons, D and B mesons, while different rows show different centrality classes. Solid black curves show the $1^{\mathrm{st}}$ order in opacity results, red dashed curves show the results up to the $2^{\mathrm{nd}}$ order, while cyan dot-dashed curves up to the $3^{\mathrm{rd}}$ order in opacity. }
\label{fig:RAA_stat}
\end{figure*}

Furthermore, Fig.~\ref{fig:RAA_el} raises another important question: as it is well known, only electric contribution exists in {\it static} QCD medium approximation~\cite{Kapusta,Le_Bellac} (though it has a different functional form compared to the electric contribution in dynamical QCD medium). That is, the magnetic contribution is inherently connected with the dynamic nature of the QCD medium. As most existing energy loss calculations assume (simplified) static QCD medium approximation, does this mean that higher orders in opacity can be neglected under such approximation?

We first note that this does not necessarily have to be the case, because the effective potential for electric contribution is significantly different in static compared to the dynamical medium. However, to address this question, we repeat the same analyses as above, this time assuming the static medium effective potential (left-hand side of Eq.~(\ref{diff_cross})) and mean free path ($\lambda_\mathrm{stat}$). Fig.~\ref{fig:RAA_stat} shows the effects of higher orders in opacity in static medium approximation. While larger than those in Fig.~\ref{fig:RAA_el}, we see that these effects are still small (i.e., less than 6\%). Thus, for optically thin medium models with static approximation, we show that including multiple scattering centers has a small effect on the numerical results, i.e., these effects can also be neglected.

\begin{figure*}
\centering
\epsfig{file=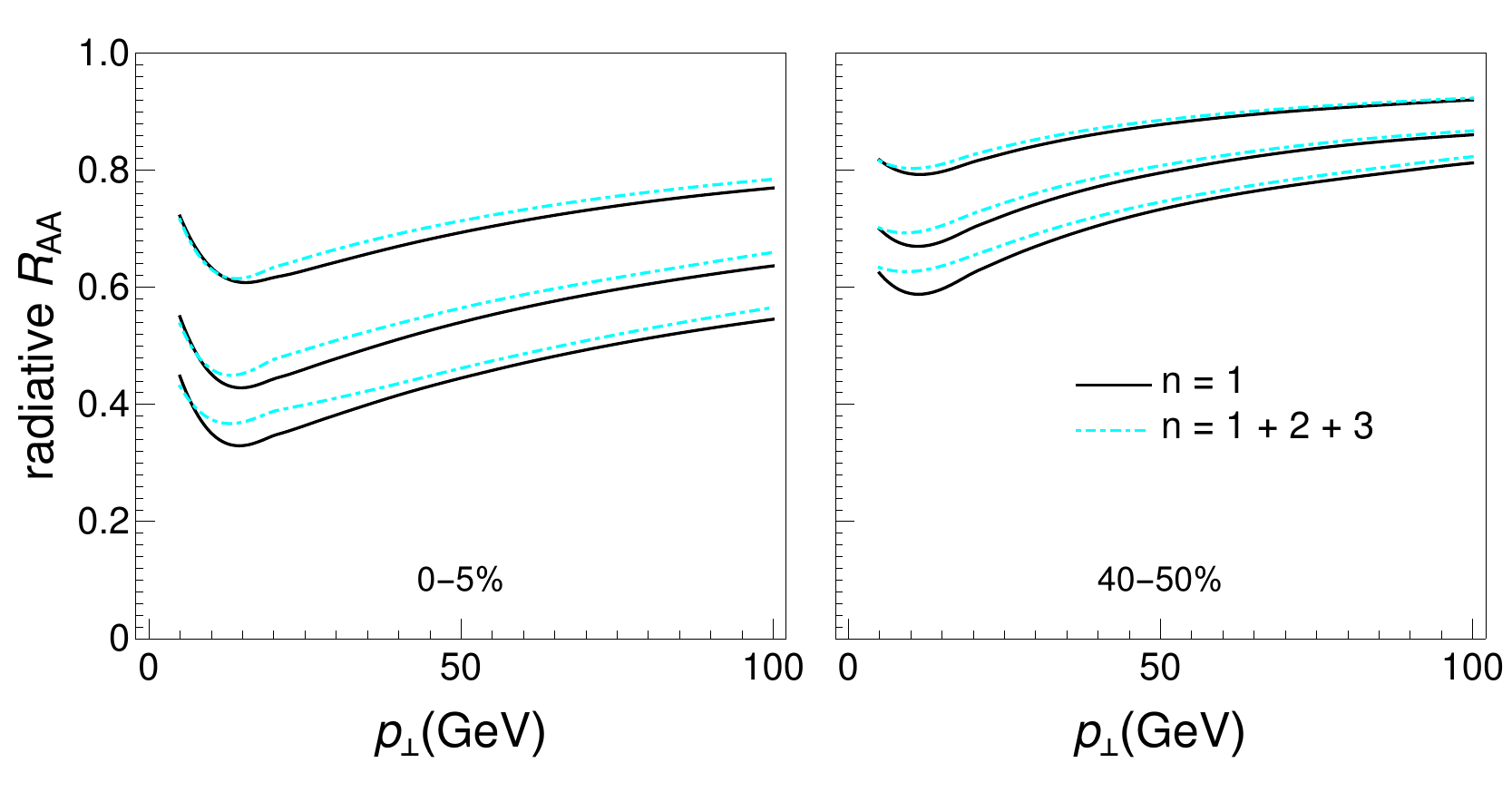,scale=0.40}
\vspace*{-0.2cm}
\caption{D meson radiative $R_{AA}$ results obtained within DREENA-C for different temperature values. The left panel corresponds to 0-5\% centrality, while the right panel corresponds to 40-50\% centrality. The values of temperature are $T=200$~MeV (the uppermost curves), $400$~MeV (the middle curves), and $600$~MeV (the lowest curves). The solid black curves show the $1^{st}$ order in opacity results, while cyan dot-dashed curves show the results up to the $3^{rd}$ order in opacity. The chromomagnetic and chromoelectric mass ratio is fixed to $\mu_M / \mu_E = 0.6$.}
\label{fig:RAA_Tdep}
\end{figure*}

Finally, we ask how the inclusion of evolving medium would modify these results. Including higher-order effects in evolving medium is very demanding and out of the scope of this manuscript. However, it can be partially addressed by studying how higher-order effects depend on the temperature, which changes in the evolving medium. To address this, in Fig.~\ref{fig:RAA_Tdep}, we focus on D meson $R_{AA}$, $\mu_M/\mu_E=0.6$ (per agreement with~\cite{Borsanyi:2015yka}) and study the effects of higher orders in opacity for three different temperature values $T=200, 400, 600$~MeV (which broadly covers the range of temperatures accessible at RHIC and the LHC). We find that the higher-order effects are largely independent of these values. Thus, we do not expect that including medium evolution will significantly influence the results presented in this study, i.e., expect the effect of multiple scattering centers to remain small.

\section{Summary}

In this manuscript, we generalized our dynamical energy loss and DGLV formalisms towards finite orders in opacity. For bottom quarks, we find that higher orders in opacity are insignificant due to short gluon formation time, i.e. the incoherent limit. For charm and light quarks, including $2^{nd}$ order in opacity is sufficient, i.e., the $3^{rd}$ order numerical results almost overlap with the $2^{nd}$. Surprisingly, we also find that for limits of magnetic screening, $\mu_M/\mu_E=0.4$ and $\mu_M/\mu_E=0.6$, the effects on the $R_{AA}$ are opposite in sign. That is, for $\mu_M/\mu_E=0.6$ ($\mu_M/\mu_E=0.4$), higher orders in opacity decrease (increase) the energy loss and subsequently suppression. The intuitive reason behind such behavior is the magnetic contribution to the dynamical energy loss. That is, while electric contribution remains almost insensitive to increases in the order of opacity, magnetic screening larger (smaller) than plasmon mass value decreases (increases) the energy loss and suppression, in agreement with the theoretical expectations. We also show that in the static QCD medium approximation, in which (per definition) only electric contribution remains, the effects of higher orders in opacity on high-$p_\perp$ observables are small and can be safely neglected. Thus, for static QCD medium, the first order in opacity is an adequate approximation for finite-size QCD medium created in the RHIC and the LHC. For dynamical energy loss, both the sign and the size of the effects depend on the magnetic screening, as outlined above. However, for most of the current estimates of magnetic screening~\cite{Borsanyi:2015yka}, these effects remain less than 5\%, so they can also be safely neglected. 

The analyses presented here are obtained for a constant temperature medium (and adequately generalized DREENA-C framework). However, we also tested how the effects of including multiple scatterers depend upon temperature, and found this influence to be also small (affecting the radiative $R_{AA}$ for less than 5\%). Thus, we expect that including higher orders in opacity in the evolving medium will not change the qualitative results obtained here, but this remains to be rigorously tested in the future.

{\em Acknowledgments:}
We thank Jussi Auvinen for the useful discussions. This work is supported by the European Research Council, grant
ERC-2016-COG: 725741, and by the Ministry of Science and Technological
Development of the Republic of Serbia. 

\appendix

\begingroup
\small

\section{Analytical expressions for $dN_g/dx$: general form}

The gluon radiation spectrum up to the $4^{th}$ order in opacity contains the following terms, which are here given in detail:

\beqar
\biggl( \frac{dN_g}{dx} \biggl) &=& \biggl( \frac{dN_g^{(1)}}{dx} \biggl) + \biggl( \frac{dN_g^{(2)}}{dx} \biggl)_{1} - \biggl( \frac{dN_g^{(2)}}{dx} \biggl)_{2} \nonumber \\
&+& \biggl( \frac{dN_g^{(3)}}{dx} \biggl)_{1} - \biggl( \frac{dN_g^{(3)}}{dx} \biggl)_{2} - \biggl( \frac{dN_g^{(3)}}{dx} \biggl)_{3} + \biggl( \frac{dN_g^{(3)}}{dx} \biggl)_{4} \nonumber \\
&+& \biggl( \frac{dN_g^{(4)}}{dx} \biggl)_{1} - \biggl( \frac{dN_g^{(4)}}{dx} \biggl)_{2} - \biggl( \frac{dN_g^{(4)}}{dx} \biggl)_{3} + \biggl( \frac{dN_g^{(4)}}{dx} \biggl)_{4} \nonumber \\
&-& \biggl( \frac{dN_g^{(4)}}{dx} \biggl)_{5} + \biggl( \frac{dN_g^{(4)}}{dx} \biggl)_{6} + \biggl( \frac{dN_g^{(4)}}{dx} \biggl)_{7} - \biggl( \frac{dN_g^{(4)}}{dx} \biggl)_{8}
\eeqar

Numerical integrations with respect to the momentum $\textbf{k}$ are performed over $0<|\textbf{k}|<2Ex(1-x)$, and the ones with respect to momenta $\textbf{q}_{i}$ are performed over $0<|\textbf{q}_{i}|<\sqrt{4ET}$~\cite{OpenCharmBeauty}. The integrations with respect to angles $\varphi_i$ are performed over $0<\varphi_i<2\pi$. Under the constant $T$ approximation considered in this manuscript, the expressions presented below can be analytically integrated over $z_i$, significantly simplifying subsequent numerical calculations (see Appendix B).

In the expressions below, the following equations hold for $i,j \in \{1,2,3,4\}$:
\beqar
\textbf{k} \cdot \textbf{q}_{i} &=& |\textbf{k}| |\textbf{q}_{i}| \cos\varphi_i, \\
\textbf{q}_{i} \cdot \textbf{q}_{j} &=& |\textbf{q}_{i}| |\textbf{q}_{j}| \cos(\varphi_i - \varphi_j).
\eeqar

The $1^{st}$ order in opacity term is given by:

\beqar
\biggl( \frac{dN_g^{(1)}}{dx} \biggl) \!&=&\! \frac{4C_{R}}{\pi x} \int_{0}^{L} dz_1
\int \frac{d^2 {\bf k}}{\pi} \int \frac{d^2{\bf q}_{1}}{\pi} \, \alpha_{s}(Q^{2}_k) \frac{1}{\lambda_{dyn}} \frac{\mu_{E}^2-\mu_{M}^2}{(\qT^2 + \mu_{E}^2)(\qT^2 + \mu_{M}^2)}
\nonumber
\\
&&\frac{\chi^2(\qT \cdot (\qT-\kT)) + (\qT \cdot \kT) (\kT-\qT)^2}{(\chi^2+\kT^2)(\chi^2+(\kT-\qT)^2)^2}
\sin^2 \biggl(\frac{\chi^2 + (\kT^2-\qT)^2}{4xE} z_1\biggl).
\eeqar

After integration with respect to $z_1$, this expression reduces to the expression used to obtain $dN_g/dx$ in the original DREENA-C framework~\cite{DREENA_C}.

The $2^{nd}$ order in opacity contains two terms, which are given by:

\beqar
\biggl( \frac{dN_g^{(2)}}{dx} \biggl)_{1}\! &=& \!\frac{4 C_{R}}{\pi x}
\int_{0}^{L} \int_{z_1}^{L} dz_1 dz_2
\int \frac{d^2 {\bf k}}{\pi}  \iint \frac{d^2{\bf q}_{1}}{\pi} \frac{d^2{\bf q}_{2}}{\pi} \nonumber \\
&& \alpha_{s}(Q^{2}_k) \frac{1}{\lambda^2_{dyn}} \frac{\mu_{E}^2-\mu_{M}^2}{(\qT^2 + \mu_{E}^2)(\qT^2 + \mu_{M}^2)}
\frac{\mu_{E}^2-\mu_{M}^2}{(\qTd^2 + \mu_{E}^2)(\qTd^2 + \mu_{M}^2)} \nonumber \\
&&\frac{\chi^2(\qTd\!\cdot\!(\qT\!+\!\qTd\!-\!\kT))\!+\!(\qTd\!\cdot\!\kT)(\kT\!-\!\qTd)^2\!+\!(\kT\!\cdot\!\qT)(\qTd\!\cdot\!(\qTd\!-\!2 \kT))\!+\!\kT^2(\qTd\!\cdot\!\qT)}
{(\chi^2 + \kT^2)(\chi^2 + (\kT - \qTd)^2)(\chi^2 + (\kT - \qT -\qTd)^2)}  \nonumber \\
&&\sin \biggl(\!\frac{\chi^2\! +\! (\kT\!-\!\qT\!-\!\qTd)^2}{4xE} z_1 \!\biggl) \sin \biggl(\!\frac{\chi^2\!+\!(\kT-\qT-\qTd)^2}{4xE} z_1 + \frac{\chi^2 + (\kT-\qTd)^2}{2xE} z_2\!\biggl),
\eeqar
\beqar
\biggl( \frac{dN_g^{(2)}}{dx} \biggl)_{2}\! &=& \!\frac{4 C_{R}}{\pi x}
\int_{0}^{L} \int_{z_1}^{L} dz_1 dz_2
\int \frac{d^2 {\bf k}}{\pi}  \int \frac{d^2{\bf q}_{2}}{\pi} \alpha_{s}(Q^{2}_k) \frac{1}{\lambda^2_{dyn}}
\frac{\mu_{E}^2-\mu_{M}^2}{(\qTd^2 + \mu_{E}^2)(\qTd^2 + \mu_{M}^2)} \nonumber \\
&& \frac{\chi^2 (\qTd \cdot (\qTd\!-\!\kT))\! +\! (\qTd \cdot \kT) (\kT-\qTd)^2}{(\chi^2+\kT^2)(\chi^2\! +\! (\kT-\qTd)^2)^2}
\sin \!\left( \frac{\chi^2 \!+\! (\kT-\qTd)^2}{4xE} z_1 \right) \sin \!\left( \frac{\chi^2 + (\kT-\qTd)^2}{2xE} (\frac{z_1}{2} \!+\! z_2) \right).
\eeqar

The $3^{rd}$ order in opacity containts four terms, which are given by:

\beqar
\biggl( \frac{dN_g^{(3)}}{dx} \biggl)_{1} \!&=&\! \frac{4C_R}{\pi x} \int_{0}^{L} \int_{z_1}^{L} \int_{z_2}^{L} dz_1 dz_2 dz_3 \int \frac{d^2 {\bf k}}{\pi}  \iiint \frac{d^2{\bf q}_{1}}{\pi} \frac{d^2{\bf q}_{2}}{\pi} \frac{d^2{\bf q}_{3}}{\pi}
\nonumber \\ &&
 \alpha_{s}(Q^{2}_k) \frac{1}{\lambda^3_{dyn}}
\frac{\mu_{E}^2-\mu_{M}^2}{(\qT^2 + \mu_{E}^2)(\qT^2 + \mu_{M}^2)}
\frac{\mu_{E}^2-\mu_{M}^2}{(\qTd^2 + \mu_{E}^2)(\qTd^2 + \mu_{M}^2)}
\frac{\mu_{E}^2-\mu_{M}^2}{(\qTt^2 + \mu_{E}^2)(\qTt^2 + \mu_{M}^2)}
\nonumber \\ &&
\frac{\chi^2(\qTt \!\cdot\! (\qT \!+\! \qTd \!+\! \qTt \!-\! \kT))\! +\! (\qTt \!\cdot\! \kT)(\kT\!-\!\qTt)^2\!
+\! (\kT \!\cdot\! (\qT\!+\!\qTd))(\qTt\!\cdot\!(\qTt\!-\!2\kT))\!+\!\kT^2(\qTt\!\cdot\!(\qT\!+\!\qTd))}{(\chi^2+\kT^2)(\chi^2+(\kT-\qTt)^2)(\chi^2+(\kT-\qT-\qTd-\qTt)^2)}
 \nonumber \\ &&   \sin \left( \frac{\chi^2 + (\kT-\qT-\qTd-\qTt)^2}{4xE} z_1  \right)
\nonumber \\ &&
    \sin \left( \frac{\chi^2 + (\kT-\qT-\qTd-\qTt)^2}{4xE}  z_1  + \frac{\chi^2 + (\kT-\qTd-\qTt)^2}{2xE} z_2 + \frac{\chi^2 + (\kT-\qTt)^2}{2xE} z_3 \right),
\eeqar

\beqar
    \biggl( \frac{dN_g^{(3)}}{dx} \biggl)_{2} \!&=&\! \frac{4C_R}{\pi x} \int_{0}^{L} \int_{z_1}^{L} \int_{z_2}^{L} dz_1 dz_2 dz_3 \int \frac{d^2 {\bf k}}{\pi}  \iint \frac{d^2{\bf q}_{1}}{\pi} \frac{d^2{\bf q}_{3}}{\pi}
\nonumber \\ &&
 \alpha_{s}(Q^{2}_k) \frac{1}{\lambda^3_{dyn}}
\frac{\mu_{E}^2-\mu_{M}^2}{(\qT^2 + \mu_{E}^2)(\qT^2 + \mu_{M}^2)} \frac{\mu_{E}^2-\mu_{M}^2}{(\qTt^2 + \mu_{E}^2)(\qTt^2 + \mu_{M}^2)}
\nonumber \\ &&
\frac{\chi^2 (\qTt \cdot (\qT+\qTt-\kT))+(\qTt\cdot\kT)(\kT-\qTt)^2+(\kT\cdot\qT)(\qTt\cdot(\qTt-2\kT))+\kT^2 (\qTt\cdot\qT)}
{(\chi^2+\kT^2)(\chi^2+(\kT-\qTt)^2)(\chi^2+(\kT-\qT-\qTt)^2)}
\nonumber \\ &&
\sin \!\left(\!\frac{\chi^2\!+\!(\kT\!-\!\qT\!-\!\qTt)^2}{4xE} z_1\! \right)
\sin \!\left(\! \frac{\chi^2\!+\!(\kT\!-\!\qT\!-\!\qTt)^2}{4xE} z_1 + \frac{\chi^2\!+\!(\kT\!-\!\qTt)^2}{2xE} (z_2 \!+ \!z_3)\! \right)\!,
\eeqar

\beqar
    \biggl( \frac{dN_g^{(3)}}{dx} \biggl)_{3} \!&=&\! \frac{4C_R}{\pi x} \int_{0}^{L} \int_{z_1}^{L} \int_{z_2}^{L} dz_1 dz_2 dz_3 \int \frac{d^2 {\bf k}}{\pi}  \iint \frac{d^2{\bf q}_{2}}{\pi} \frac{d^2{\bf q}_{3}}{\pi}
\nonumber \\ &&
\alpha_{s}(Q^{2}_k) \frac{1}{\lambda^3_{dyn}}
\frac{\mu_{E}^2-\mu_{M}^2}{(\qTd^2 + \mu_{E}^2)(\qTd^2 + \mu_{M}^2)}
\frac{\mu_{E}^2-\mu_{M}^2}{(\qTt^2 + \mu_{E}^2)(\qTt^2 + \mu_{M}^2)}
\nonumber \\ &&
\frac{\chi^2(\qTt\cdot(\qTd+\qTt-\kT))+(\qTt\cdot\kT)(\kT-\qTt)^2+(\kT\cdot\qTd)(\qTt\cdot(\qTt-2\kT))+\kT^2(\qTt\cdot\qTd)}{(\chi^2+\kT^2)(\chi^2+(\kT-\qTt)^2)(\chi^2+(\kT-\qTd-\qTt)^2)} \nonumber \\ &&
\sin\! \left(\! \frac{\chi^2\! + \!(\kT\!-\!\qTd\!-\!\qTt)^2}{4xE} z_1\! \right)
\sin \!\left( \!\frac{\chi^2\!+\!(\kT\!-\!\qTd\!-\!\qTt)^2}{2xE}\! \left( \!\frac{z_1}{2}\! +\! z_2 \!\right) \!+\!
\frac{\chi^2\!+\!(\kT\!-\!\qTt)^2}{2xE} z_3 \!\right)\!,
\eeqar

\beqar
    \biggl( \frac{dN_g^{(3)}}{dx} \biggl)_{4} \!&=&\! \frac{4C_R}{\pi x} \int_{0}^{L} \int_{z_1}^{L} \int_{z_2}^{L} dz_1 dz_2 dz_3 \int \frac{d^2 {\bf k}}{\pi}
\int \frac{d^2{\bf q}_{3}}{\pi} \nonumber \\ &&
\alpha_{s}(Q^{2}_k) \frac{1}{\lambda^3_{dyn}}
\frac{\mu_{E}^2-\mu_{M}^2}{(\qTt^2 + \mu_{E}^2)(\qTt^2 + \mu_{M}^2)} \frac{\chi^2(\qTt\cdot(\qTt-\kT))+(\qTt\cdot\kT)(\kT-\qTt)^2}{(\chi^2+\kT^2)(\chi^2+(\kT-\qTt)^2)^2}
\nonumber \\ &&
\sin \left( \frac{\chi^2+(\kT-\qTt)^2}{4xE} z_1 \right)
\sin \left( \frac{\chi^2 + (\kT-\qTt)^2}{2xE} \left(\frac{z_1}{2}+z_2+z_3 \right) \right).
\eeqar

The $4^{th}$ order in opacity is given by eight terms, which are given by:

\beqar
\biggl( \frac{dN_g^{(4)}}{dx} \biggl)_{1} \!&=&\! \frac{4C_R}{\pi x} \int_{0}^{L} \int_{z_1}^{L} \int_{z_2}^{L} \int_{z_3}^{L} dz_1 dz_2 dz_3 dz_4 \int \frac{d^2 {\bf k}}{\pi}
\iiiint \frac{d^2{\bf q}_{1}}{\pi} \frac{d^2{\bf q}_{2}}{\pi} \frac{d^2{\bf q}_{3}}{\pi} \frac{d^2{\bf q}_{4}}{\pi}
\nonumber \\ &&
 \alpha_{s}(Q^{2}_k) \frac{1}{\lambda^4_{dyn}}
\frac{\mu_{E}^2-\mu_{M}^2}{(\qT^2 + \mu_{E}^2)(\qT^2 + \mu_{M}^2)}
\frac{\mu_{E}^2-\mu_{M}^2}{(\qTd^2 + \mu_{E}^2)(\qTd^2 + \mu_{M}^2)}
\frac{\mu_{E}^2-\mu_{M}^2}{(\qTt^2 + \mu_{E}^2)(\qTt^2 + \mu_{M}^2)}
\frac{\mu_{E}^2-\mu_{M}^2}{(\qTc^2 + \mu_{E}^2)(\qTc^2 + \mu_{M}^2)}
\nonumber \\ &&
\frac{\chi^2 ( \!\qTc \!\cdot\! (\qT\!+\!\qTd\!+\!\qTt\!+\!\qTc\!-\!\kT))\!+\!(\qTc\!\cdot\!\kT)(\kT\!-\!\qTc)^2
\!+\!(\kT\!\cdot\!(\qT\!+\!\qTd\!+\!\qTt))(\qTc\!\cdot\!(\qTc\!-\!2\kT))\!+\!\kT^2(\qTc\!\cdot\!(\qT\!+\!\qTd\!+\!\qTt))}{(\chi^2\!+\!\kT^2)(\chi^2\!+\!(\kT\!-\!\qTc)^2)
(\chi^2\!+\!(\kT\!-\!\qT\!-\!\qTd\!-\!\qTt\!-\!\qTc)^2)}
\nonumber \\ &&
\sin \left( \!\frac{\chi^2\!+\!(\kT\!-\!\qT\!-\!\qTd\!-\!\qTt\!-\!\qTc)^2}{4xE} z_1 \!+\! \frac{\chi^2\!+\!(\kT\!-\!\qTd\!-\!\qTt\!-\!\qTc)^2}{2xE} z_2 \!+\!
\frac{\chi^2\!+\!(\kT\!-\!\qTt\!-\!\qTc)^2}{2xE} z_3 \!+\! \frac{\chi^2\!+\!(\kT\!-\!\qTc)^2}{2xE} z_4\! \right)
\nonumber \\ &&
\sin \left( \frac{\chi^2+(\kT-\qT-\qTd-\qTt-\qTc)^2}{4xE} z_1 \right),
\eeqar

\beqar
\biggl( \!\frac{dN_g^{(4)}}{dx}\! \biggl)_{2}\!&=&\!\frac{4C_R}{\pi x} \int_{0}^{L} \int_{z_1}^{L} \int_{z_2}^{L} \int_{z_3}^{L} dz_1 dz_2 dz_3 dz_4 \int \frac{d^2 {\bf k}}{\pi} \iiint \frac{d^2{\bf q}_{1}}{\pi} \frac{d^2{\bf q}_{2}}{\pi} \frac{d^2{\bf q}_{4}}{\pi}
\nonumber \\ &&
\alpha_{s}(Q^{2}_k) \frac{1}{\lambda^4_{dyn}}
\frac{\mu_{E}^2-\mu_{M}^2}{(\qT^2 + \mu_{E}^2)(\qT^2 + \mu_{M}^2)}
\frac{\mu_{E}^2-\mu_{M}^2}{(\qTd^2 + \mu_{E}^2)(\qTd^2 + \mu_{M}^2)}
\frac{\mu_{E}^2-\mu_{M}^2}{(\qTc^2 + \mu_{E}^2)(\qTc^2 + \mu_{M}^2)} \nonumber\\ &&
\frac{\chi^2 (\qTc\cdot(\qT+\qTd+\qTc-\kT))+(\qTc\cdot\kT)(\kT-\qTc)^2 + (\kT\cdot(\qT+\qTd))(\qTc\cdot(\qTc-2\kT))+\kT^2(\qTc\cdot(\qT+\qTd)) }{(\chi^2+\kT^2)(\chi^2+(\kT-\qTc)^2)(\chi^2+(\kT-\qT-\qTd-\qTc)^2)}
\nonumber \\ &&
\sin \! \left(\! \frac{\chi^2\!+\!(\kT\!-\!\qT\!-\!\qTd\!-\!\qTc)^2}{4xE}\! z_1 \!+\!
\frac{\chi^2\!+\!(\kT\!-\!\qTd\!-\!\qTc)^2}{2xE} z_2 \!+\! \frac{\chi^2\!+\!(\kT\!-\!\qTc)^2}{2xE} \! (z_3\!+\!z_4)\! \right)
\nonumber \\ &&
\sin \!\left( \!\frac{\chi^2\!+\!(\kT\!-\!\qT\!-\!\qTd\!-\!\qTc)^2}{4xE} z_1 \!\right),
\eeqar

\beqar
\biggl( \frac{dN_g^{(4)}}{dx} \biggl)_{3} \!&=&\!  \frac{4C_R}{\pi x} \int_{0}^{L} \int_{z_1}^{L} \int_{z_2}^{L} \int_{z_3}^{L} dz_1 dz_2 dz_3 dz_4 \int \frac{d^2 {\bf k}}{\pi}  \iiint \frac{d^2{\bf q}_{1}}{\pi} \frac{d^2{\bf q}_{3}}{\pi} \frac{d^2{\bf q}_{4}}{\pi}
\nonumber \\ &&
 \alpha_{s}(Q^{2}_k) \frac{1}{\lambda^4_{dyn}}
\frac{\mu_{E}^2-\mu_{M}^2}{(\qT^2 + \mu_{E}^2)(\qT^2 + \mu_{M}^2)}
\frac{\mu_{E}^2-\mu_{M}^2}{(\qTt^2 + \mu_{E}^2)(\qTt^2 + \mu_{M}^2)}
\frac{\mu_{E}^2-\mu_{M}^2}{(\qTc^2 + \mu_{E}^2)(\qTc^2 + \mu_{M}^2)}\nonumber \\ &&
\frac{\chi^2(\qTc\cdot(\qT+\qTt+\qTc-\kT))+(\qTc\cdot\kT)(\kT-\qTc)^2
+(\kT\cdot(\qT+\qTt))(\qTc\cdot(\qTc-2\kT))+\kT^2(\qTc\cdot(\qT+\qTt))}{(\chi^2+\kT^2)(\chi^2+(\kT-\qTc)^2)(\chi^2+(\kT-\qT-\qTt-\qTc)^2)}
\nonumber \\ &&
\sin \left( \frac{\chi^2+(\kT-\qT-\qTt-\qTc)^2}{4xE} z_1 +
\frac{\chi^2 + (\kT-\qTt-\qTc)^2}{2xE} (z_2+z_3) +
\frac{\chi^2 + (\kT-\qTc)^2}{2xE} z_4 \right)
\nonumber \\ &&
\sin \left( \frac{\chi^2+(\kT-\qT-\qTt-\qTc)^2}{4xE} z_1 \right),
\eeqar

\beqar
\biggl( \frac{dN_g^{(4)}}{dx} \biggl)_{4} \!&=&\!  \frac{4C_R}{\pi x} \int_{0}^{L} \int_{z_1}^{L} \int_{z_2}^{L} \int_{z_3}^{L} dz_1 dz_2 dz_3 dz_4 \int \frac{d^2 {\bf k}}{\pi}  \iint \frac{d^2{\bf q}_{1}}{\pi} \frac{d^2{\bf q}_{4}}{\pi}
\nonumber \\ &&
 \alpha_{s}(Q^{2}_k) \frac{1}{\lambda^4_{dyn}}
\frac{\mu_{E}^2-\mu_{M}^2}{(\qT^2 + \mu_{E}^2)(\qT^2 + \mu_{M}^2)}
\frac{\mu_{E}^2-\mu_{M}^2}{(\qTc^2 + \mu_{E}^2)(\qTc^2 + \mu_{M}^2)}
\nonumber \\ &&
\frac{\chi^2(\qTc\cdot(\qT+\qTc-\kT))+(\qTc\cdot\kT)(\kT-\qTc)^2+(\kT\cdot\qT)(\qTc\cdot(\qTc-2\kT))+\kT^2(\qTc\cdot\qT) }{(\chi^2+\kT^2)(\chi^2+(\kT-\qTc)^2)(\chi^2+(\kT-\qT-\qTc)^2)}
\nonumber \\ &&
\sin \left(\! \frac{\chi^2\!+\!(\kT\!-\!\qT\!-\!\qTc)^2}{4xE} z_1 \!\right)
\sin \left(\! \frac{\chi^2\!+\!(\kT\!-\!\qT\!-\!\qTc)^2}{4xE} z_1 \!+\!
\frac{\chi^2 \!+\! (\kT\!-\!\qTc)^2}{2xE} (z_2 \!+\! z_3 \!+ \!z_4)\! \right),
\eeqar

\beqar
\biggl( \frac{dN_g^{(4)}}{dx} \biggl)_{5} \!&=&\!  \frac{4C_R}{\pi x} \int_{0}^{L} \int_{z_1}^{L} \int_{z_2}^{L} \int_{z_3}^{L} dz_1 dz_2 dz_3 dz_4 \int \frac{d^2 {\bf k}}{\pi}  \iiint \frac{d^2{\bf q}_{2}}{\pi} \frac{d^2{\bf q}_{3}}{\pi} \frac{d^2{\bf q}_{4}}{\pi}
\nonumber \\ &&
 \alpha_{s}(Q^{2}_k) \frac{1}{\lambda^4_{dyn}}
\frac{\mu_{E}^2-\mu_{M}^2}{(\qTd^2 + \mu_{E}^2)(\qTd^2 + \mu_{M}^2)}
\frac{\mu_{E}^2-\mu_{M}^2}{(\qTt^2 + \mu_{E}^2)(\qTt^2 + \mu_{M}^2)}
\frac{\mu_{E}^2-\mu_{M}^2}{(\qTc^2 + \mu_{E}^2)(\qTc^2 + \mu_{M}^2)}
\nonumber \\ &&
\frac{\chi^2(\qTc\cdot(\qTd+\qTt+\qTc-\kT))+(\qTc\cdot\kT)(\kT-\qTc)^2
+(\kT\cdot(\qTd+\qTt))(\qTc\cdot(\qTc-2\kT))+\kT^2(\qTc\cdot(\qTd+\qTt))}{(\chi^2+\kT^2)(\chi^2+(\kT-\qTc)^2)(\chi^2+(\kT-\qTd-\qTt-\qTc)^2)}
\nonumber \\ &&
\sin \left( \frac{\chi^2+(\kT-\qTd-\qTt-\qTc)^2}{2xE} (\frac{z_1}{2} + z_2) +
\frac{\chi^2 + (\kT-\qTt-\qTc)^2}{2xE} z_3 +
\frac{\chi^2 + (\kT-\qTc)^2}{2xE} z_4 \right)
\nonumber \\ &&
\sin \left( \frac{\chi^2+(\kT-\qTd-\qTt-\qTc)^2}{4xE} z_1 \right),
\eeqar

\beqar
\biggl( \frac{dN_g^{(4)}}{dx} \biggl)_{6} \!&=&\! \frac{4C_R}{\pi x} \int_{0}^{L} \int_{z_1}^{L} \int_{z_2}^{L} \int_{z_3}^{L} dz_1 dz_2 dz_3 dz_4 \int \frac{d^2 {\bf k}}{\pi}  \iint \frac{d^2{\bf q}_{2}}{\pi} \frac{d^2{\bf q}_{4}}{\pi}
\nonumber \\ &&
 \alpha_{s}(Q^{2}_k) \frac{1}{\lambda^4_{dyn}}
\frac{\mu_{E}^2-\mu_{M}^2}{(\qTd^2 + \mu_{E}^2)(\qTd^2 + \mu_{M}^2)}
\frac{\mu_{E}^2-\mu_{M}^2}{(\qTc^2 + \mu_{E}^2)(\qTc^2 + \mu_{M}^2)}
\nonumber \\ &&
\frac{\chi^2(\qTc\cdot(\qTd+\qTc-\kT))+(\qTc\cdot\kT)(\kT-\qTc)^2+(\kT\cdot\qTd)(\qTc\cdot(\qTc-2\kT))+\kT^2(\qTc\cdot\qTd)  }{(\chi^2+\kT^2)(\chi^2+(\kT-\qTc)^2)(\chi^2+(\kT-\qTd-\qTc)^2)}
\nonumber \\ &&
\sin \left(\! \frac{\chi^2\!+\!(\kT\!-\!\qTd\!-\!\qTc)^2}{4xE} z_1 \!\right)
\sin \left(\! \frac{\chi^2\!+\!(\kT\!-\!\qTd\!-\!\qTc)^2}{2xE} (\frac{z_1}{2} \!+\! z_2) \!+\!
\frac{\chi^2 \!+\! (\kT\!-\!\qTc)^2}{2xE} (z_3 \!+\! z_4) \!\right),
\eeqar

\beqar
\biggl( \frac{dN_g^{(4)}}{dx} \biggl)_{7} \!&=&\!  \frac{4C_R}{\pi x} \int_{0}^{L} \int_{z_1}^{L} \int_{z_2}^{L} \int_{z_3}^{L} dz_1 dz_2 dz_3 dz_4 \int \frac{d^2 {\bf k}}{\pi}  \iint \frac{d^2{\bf q}_{3}}{\pi} \frac{d^2{\bf q}_{4}}{\pi}
\nonumber \\ &&
 \alpha_{s}(Q^{2}_k) \frac{1}{\lambda^4_{dyn}}
\frac{\mu_{E}^2-\mu_{M}^2}{(\qTt^2 + \mu_{E}^2)(\qTt^2 + \mu_{M}^2)}
\frac{\mu_{E}^2-\mu_{M}^2}{(\qTc^2 + \mu_{E}^2)(\qTc^2 + \mu_{M}^2)}  \nonumber \\ &&
\frac{\chi^2(\qTc\cdot(\qTt+\qTc-\kT))+(\qTc\cdot\kT)(\kT-\qTc)^2 + (\kT\cdot\qTt)(\qTc\cdot(\qTc-2\kT))+\kT^2(\qTc\cdot\qTt)}{(\chi^2+\kT^2)(\chi^2+(\kT-\qTc)^2)(\chi^2+(\kT-\qTt-\qTc)^2)}
\nonumber \\ &&
\sin \left(\! \frac{\chi^2\!+\!(\kT\!-\!\qTt\!-\!\qTc)^2}{4xE} z_1 \!\right)
\sin \left(\! \frac{\chi^2\!+\!(\kT\!-\!\qTt\!-\!\qTc)^2}{2xE} (\frac{z_1}{2} \!+\! z_2 \!+\! z_3) \!+\! \frac{\chi^2 \!+\! (\kT\!-\!\qTc)^2}{2xE} z_4 \!\right),
\eeqar

\beqar
\biggl( \frac{dN_g^{(4)}}{dx} \biggl)_{8} \!&=&\!  \frac{4C_R}{\pi x} \int_{0}^{L} \int_{z_1}^{L} \int_{z_2}^{L} \int_{z_3}^{L} dz_1 dz_2 dz_3 dz_4 \int \frac{d^2 {\bf k}}{\pi}  \int \frac{d^2{\bf q}_{4}}{\pi}
\nonumber \\ &&
 \alpha_{s}(Q^{2}_k) \frac{1}{\lambda^4_{dyn}}
\frac{\mu_{E}^2-\mu_{M}^2}{(\qTc^2 + \mu_{E}^2)(\qTc^2 + \mu_{M}^2)}
\frac{\chi^2(\qTc\cdot(\qTc-\kT)) + (\qTc\cdot\kT)(\kT-\qTc)^2}{(\chi^2+\kT^2)(\chi^2+(\kT-\qTc)^2)^2}
\nonumber \\ &&
\sin \left( \frac{\chi^2 + (\kT-\qTc)^2}{4xE} z_1 \right)
\sin \left( \frac{\chi^2 + (\kT-\qTc)^2}{2xE} (\frac{z_1}{2} + z_2 + z_3 + z_4) \right).
\eeqar

\section{Analytical expressions for $dN_g/dx$ within DREENA-C}

Within the DREENA-C framework, under the assumption of constant medium temperature, we can explicitly perform analytical integrations for $z_i$, where $(i=1,2,3,4)$. $\omega_{(m \ldots n)}$ coefficients are defined in the Theoretical framework section. The expression for the $1^{st}$ order in opacity then became:

\beqar
\biggl( \frac{dN_g^{(1)}}{dx} \biggl) \!&=&\!  \frac{2 C_{R}}{\pi x}
\int \frac{d^2 {\bf k}}{\pi}  \int \frac{d^2{\bf q}_{1}}{\pi} \,
\alpha_{s}(Q^{2}_k) \frac{L}{\lambda_{dyn}}
\frac{\mu_{E}^2-\mu_{M}^2}{(\qT^2 + \mu_{E}^2)(\qT^2 + \mu_{M}^2)}
\nonumber \\ &&
\frac{\chi^2(\qT \cdot (\qT-\kT)) + (\qT \cdot \kT) (\kT-\qT)^2}{(\chi^2+\kT^2)(\chi^2+(\kT-\qT)^2)^2}
\left(1 - \frac{\sin(L\omega_{(1)})}{L \omega_{(1)}} \right) ,
\eeqar
The expressions for higher orders in opacity became:
\beqar
\biggl( \!\frac{dN_g^{(2)}\!}{dx} \biggl)_{1} \!&=&\!  \frac{2 C_{R}}{\pi x}
\int \frac{d^2 {\bf k}}{\pi} \iint \frac{d^2{\bf q}_{1}}{\pi} \frac{d^2{\bf q}_{2}}{\pi} 
\, \alpha_{s}(Q^{2}_k) \frac{1}{\lambda^2_{dyn}} \frac{\mu_{E}^2-\mu_{M}^2}{(\qT^2 + \mu_{E}^2)(\qT^2 + \mu_{M}^2)}
\frac{\mu_{E}^2-\mu_{M}^2}{(\qTd^2 + \mu_{E}^2)(\qTd^2 + \mu_{M}^2)} \nonumber \\ &&
\frac{\chi^2 (\qTd \cdot (\qT+\qTd-\kT))+(\qTd \cdot \kT)(\kT-\qTd)^2 + (\kT \cdot \qT) (\qTd \cdot (\qTd - 2 \kT)) + \kT^2(\qTd \cdot \qT)}{(\chi^2 + \kT^2)(\chi^2 + (\kT - \qTd)^2)(\chi^2 + (\kT - \qT -\qTd)^2)}
\nonumber \\ &&
\frac{1}{\omega_{(2)}} \!\left( \!\frac{\omega _{(2)} \cos \!\left(L \! \left(\omega_{(2)}\!+\!\omega_{(12)}\!\right)\right)}{\left(\omega_{(2)}\!+\!\omega_{(12)}\right)
   \omega_{(12)}}\!+\!L \sin \!\left(L \omega_{(2)}\right)\!-\!\frac{\left(\omega_{(2)}\!-\!\omega_{(12)}\right) \cos \!\left(L \omega_{(2)}\right)}{\omega_{(2)} \omega_{(12)}}\!-\!\frac{\omega_{(12)}}{\omega_{(2)} \! \left(\omega_{(2)}\!+\!\omega_{(12)}\right)}\! \right),
\eeqar
\beqar
\biggl( \frac{dN_g^{(2)}}{dx} \biggl)_{2} \!&=&\!  \frac{2 C_{R}}{\pi x}
\int \frac{d^2 {\bf k}}{\pi} \int \frac{d^2{\bf q}_{2}}{\pi}\, \alpha_{s}(Q^{2}_k) \frac{1}{\lambda^2_{dyn}}
\frac{\mu_{E}^2-\mu_{M}^2}{(\qTd^2 + \mu_{E}^2)(\qTd^2 + \mu_{M}^2)}
\nonumber \\ &&
     \frac{\chi^2 (\qTd \cdot (\qTd-\kT)) + (\qTd \cdot \kT) (\kT-\qTd)^2}{(\chi^2+\kT^2)(\chi^2 + (\kT-\qTd)^2)^2}
\frac{\sin \left(L \omega_{(2)}\right) \left(L \omega_{(2)}-\sin \left(L \omega_{(2)}\right)\right)}{\omega_{(2)}^2},
\eeqar
\beqar
\biggl( \frac{dN_g^{(3)}}{dx} \biggl)_{1} \!&=&\! \frac{2C_R}{\pi x} \int \frac{d^2 {\bf k}}{\pi}  \iiint \frac{d^2{\bf q}_{1}}{\pi} \frac{d^2{\bf q}_{2}}{\pi} \frac{d^2{\bf q}_{3}}{\pi}
\nonumber \\ &&
 \alpha_{s}(Q^{2}_k) \frac{1}{\lambda^3_{dyn}}
\frac{\mu_{E}^2-\mu_{M}^2}{(\qT^2 + \mu_{E}^2)(\qT^2 + \mu_{M}^2)}
\frac{\mu_{E}^2-\mu_{M}^2}{(\qTd^2 + \mu_{E}^2)(\qTd^2 + \mu_{M}^2)}
\frac{\mu_{E}^2-\mu_{M}^2}{(\qTt^2 + \mu_{E}^2)(\qTt^2 + \mu_{M}^2)}
\nonumber \\ &&
\frac{\chi^2(\qTt \!\cdot\! (\qT \!+\! \qTd \!+\! \qTt \!-\! \kT))\! +\! (\qTt \!\cdot\! \kT)(\kT\!-\!\qTt)^2\!
+\! (\kT \!\cdot\! (\qT\!+\!\qTd))(\qTt\!\cdot\!(\qTt\!-\!2\kT))\!+\!\kT^2(\qTt\!\cdot\!(\qT\!+\!\qTd))}{(\chi^2+\kT^2)(\chi^2+(\kT-\qTt)^2)(\chi^2+(\kT-\qT-\qTd-\qTt)^2)}
\nonumber \\ &&
    \bigg(
	\frac{\omega_{(3)} \omega_{(123)} + 2\omega_{(23)}\omega_{(123)}-\omega_{(23)}^2 - \omega_{(3)}\omega_{(23)}}{\omega_{(23)}^2 (\omega_{(3)}+\omega_{(23)})^2 \omega_{(123)}} \sin \left(L(\omega_{(3)}+\omega_{(23)})\right)
    -\frac{\omega_{(123)} \sin \left(L
   \omega_{(3)}\right)}{\omega_{(3)} \omega_{(23)}^2 \left(\omega_{(23)}+\omega_{(123)}\right)}
\nonumber \\ &&
+\frac{\sin \left(L \left(\omega
   _{(3)}+\omega_{(23)}+\omega_{(123)}\right)\right)}{\omega_{(123)} \left(\omega_{(23)}+\omega_{(123)}\right) \left(\omega_{(3)}+\omega
   _{(23)}+\omega_{(123)}\right)}-\frac{L \cos \left(L \left(\omega_{(3)}+\omega_{(23)}\right)\right)}{\omega_{(23)} \left(\omega_{(3)}+\omega_{(23)}\right)} \bigg),
\eeqar
\beqar
    \biggl( \!\frac{dN_g^{(3)}}{dx} \biggl)_{2} \!&=&\!  \frac{C_R}{\pi x} \int \frac{d^2 {\bf k}}{\pi}  \iint \frac{d^2{\bf q}_{1}}{\pi} \frac{d^2{\bf q}_{3}}{\pi}
 \alpha_{s}(Q^{2}_k) \frac{1}{\lambda^3_{dyn}}
\frac{\mu_{E}^2-\mu_{M}^2}{(\qT^2 + \mu_{E}^2)(\qT^2 + \mu_{M}^2)} \frac{\mu_{E}^2-\mu_{M}^2}{(\qTt^2 + \mu_{E}^2)(\qTt^2 + \mu_{M}^2)}
\nonumber \\ &&
\frac{\chi^2 (\qTt \cdot (\qT+\qTt-\kT))+(\qTt\cdot\kT)(\kT-\qTt)^2+(\kT\cdot\qT)(\qTt\cdot(\qTt-2\kT))+\kT^2(\qTt\cdot\qT)}
{(\chi^2+\kT^2)(\chi^2+(\kT-\qTt)^2)(\chi^2+(\kT-\qT-\qTt)^2)}
\nonumber \\ &&
\left(\! \frac{\left(\!\frac{3\omega_{(13)}}{2}\!-\!\omega_{(3)}\!\right) \sin\! \left(2 L \omega_{(3)}\right)}{\omega_{(3)}^3 \omega_{(13)}}
\!-\!\frac{2\omega_{(13)} \sin\! \left(\!L \omega_{(3)}\!\right)}{\omega_{(3)}^3 \!\left(\omega_{(3)}\!+\!\omega_{(13)}\right)}
\!+\!\frac{\sin \!\left(2 L \left(\omega_{(3)}\!+\!\frac{\omega_{(13)}}{2}\!\right)\!\right)}{\left(\omega_{(3)}\!+\!\frac{\omega_{(13)}}{2}\!\right) \omega
   _{(13)} \left(\omega_{(3)}\!+\!\omega_{(13)}\right)}\!-\!\frac{L \!\cos \left(2 L \omega_{(3)}\!\right)}{\omega_{(3)}^2} \!\right)\!,\nonumber \\ &&
\eeqar
\beqar
    \biggl( \frac{dN_g^{(3)}}{dx} \biggl)_{3} \!&=&\!  \frac{2C_R}{\pi x} \int \frac{d^2 {\bf k}}{\pi}  \iint \frac{d^2{\bf q}_{2}}{\pi} \frac{d^2{\bf q}_{3}}{\pi} 
\alpha_{s}(Q^{2}_k) \frac{1}{\lambda^3_{dyn}}
\frac{\mu_{E}^2-\mu_{M}^2}{(\qTd^2 + \mu_{E}^2)(\qTd^2 + \mu_{M}^2)}
\frac{\mu_{E}^2-\mu_{M}^2}{(\qTt^2 + \mu_{E}^2)(\qTt^2 + \mu_{M}^2)} \nonumber \\ &&
\frac{\chi^2(\qTt\cdot(\qTd+\qTt-\kT))+(\qTt\cdot\kT)(\kT-\qTt)^2+(\kT\cdot\qTd)(\qTt\cdot(\qTt-2\kT))+\kT^2(\qTt\cdot\qTd)}{(\chi^2+\kT^2)(\chi^2+(\kT-\qTt)^2)(\chi^2+(\kT-\qTd-\qTt)^2)} \nonumber \\ &&
\left(\!\frac{\sin \left(2 L \left(\frac{\omega_{(3)}}{2}+\omega
   _{(23)}\right)\right)}{4 \omega_{(23)}^2 \left(\frac{\omega_{(3)}}{2}+\omega_{(23)}\right)}
   \!-\!\frac{\sin \left(L \omega_{(3)}\right)}{2\omega_{(23)}^2 \omega_{(3)}}\!+\!
   \frac{\frac{\sin \left(L\left(\omega_{(3)}+\omega_{(23)}\right)\right)}{\omega_{(3)}+\omega_{(23)}}\!-\!L \cos \left(L \left(\omega_{(3)}+\omega
   _{(23)}\right)\right)}{\omega_{(23)}\left(\omega_{(3)}+\omega_{(23)}\right)} \!\right)\!,
\eeqar
\beqar
    \biggl( \frac{dN_g^{(3)}}{dx} \biggl)_{4} \!&=&\! \frac{C_R}{\pi x} \int \frac{d^2 {\bf k}}{\pi}
\int \frac{d^2{\bf q}_{3}}{\pi}
\alpha_{s}(Q^{2}_k) \frac{1}{\lambda^3_{dyn}}
\frac{\mu_{E}^2-\mu_{M}^2}{(\qTt^2 + \mu_{E}^2)(\qTt^2 + \mu_{M}^2)} \frac{\chi^2(\qTt\cdot(\qTt-\kT))+(\qTt\cdot\kT)(\kT-\qTt)^2}{(\chi^2+\kT^2)(\chi^2+(\kT-\qTt)^2)^2} \nonumber \\ &&
\frac{1}{\omega_{(3)}^2} \left( -\frac{\sin \left(L \omega_{(3)}\right)}{\omega_{(3)}}+\frac{\sin \left(2 L \omega_{(3)}\right)}{2 \omega_{(3)}}+\frac{\sin \left(3 L \omega_{(3)}\right)}{3 \omega_{(3)}}-L \cos \left(2 L \omega_{(3)}\right) \right),
\eeqar
\beqar
\biggl( \frac{dN_g^{(4)}}{dx} \biggl)_{1} \!&=&\! \frac{2C_R}{\pi x} \int \frac{d^2 {\bf k}}{\pi}
\iiiint \frac{d^2{\bf q}_{1}}{\pi} \frac{d^2{\bf q}_{2}}{\pi} \frac{d^2{\bf q}_{3}}{\pi} \frac{d^2{\bf q}_{4}}{\pi}
\nonumber \\ &&
 \alpha_{s}(Q^{2}_k) \frac{1}{\lambda^4_{dyn}}
\frac{\mu_{E}^2-\mu_{M}^2}{(\qT^2 + \mu_{E}^2)(\qT^2 + \mu_{M}^2)}
\frac{\mu_{E}^2-\mu_{M}^2}{(\qTd^2 + \mu_{E}^2)(\qTd^2 + \mu_{M}^2)}
\frac{\mu_{E}^2-\mu_{M}^2}{(\qTt^2 + \mu_{E}^2)(\qTt^2 + \mu_{M}^2)}
\frac{\mu_{E}^2-\mu_{M}^2}{(\qTc^2 + \mu_{E}^2)(\qTc^2 + \mu_{M}^2)}
\nonumber \\ &&
\frac{\chi^2 ( \!\qTc \!\cdot\! (\qT\!+\!\qTd\!+\!\qTt\!+\!\qTc\!-\!\kT))\!+\!(\qTc\!\cdot\!\kT)(\kT\!-\!\qTc)^2
\!+\!(\kT\!\cdot\!(\qT\!+\!\qTd\!+\!\qTt))(\qTc\!\cdot\!(\qTc\!-\!2\kT))\!+\!\kT^2(\qTc\!\cdot\!(\qT\!+\!\qTd\!+\!\qTt))}{(\chi^2\!+\!\kT^2)(\chi^2\!+\!(\kT\!-\!\qTc)^2)
(\chi^2\!+\!(\kT\!-\!\qT\!-\!\qTd\!-\!\qTt\!-\!\qTc)^2)}
\nonumber \\ &&
\bigg(-\frac{L \sin \left(L \left(\omega_{(4)}+\omega_{(34)}+\omega_{(234)}\right)\right)}{\omega_{(234)} \left(\omega_{(34)}+\omega_{(234)}\right) \left(\omega_{(4)}+\omega_{(34)}+\omega_{(234)}\right)}\nonumber \\ &&
-\frac{\cos \left(L \left(\omega_{(4)}+\omega_{(34)}+\omega_{(234)}+\omega_{(1234)}\right)\right)}{\omega_{(1234)} \left(\omega_{(234)}+\omega_{(1234)}\right) \left(\omega_{(34)}+\omega_{(234)}+\omega_{(1234)}\right) \left(\omega_{(4)}+\omega_{(34)}+\omega_{(234)}+\omega_{(1234)}\right)}
\nonumber \\ &&
   +\frac{F_{41}}{\omega_{(234)}^2 \left(\omega_{(34)}+\omega_{(234)}\right){}^2 \left(\omega_{(4)}+\omega_{(34)}+\omega_{(234)}\right){}^2 \omega_{(1234)}} \cos \left(L(\omega_{(4)}+\omega_{(34)}+\omega_{(234)}) \right)\nonumber \\ &&
+\frac{\omega_{(1234)} \cos \left(L \left(\omega_{(4)}+\omega_{(34)}\right)\right)}{\omega_{(34)} \left(\omega_{(4)}+\omega_{(34)}\right) \omega_{(234)}^2 \left(\omega_{(234)}+\omega_{(1234)}\right)}
   -\frac{\omega_{(1234)} \cos \left(L \omega_{(4)}\right)}{\omega_{(4)} \omega_{(34)} \left(\omega_{(34)}+\omega_{(234)}\right){}^2 \left(\omega_{(34)}+\omega_{(234)}+\omega_{(1234)}\right)}\nonumber \\ &&
+\frac{\omega_{(1234)}}{\omega_{(4)} \left(\omega_{(4)}+\omega_{(34)}\right) \left(\omega_{(4)}+\omega_{(34)}+\omega_{(234)}\right){}^2 \left(\omega_{(4)}+\omega_{(34)}+\omega_{(234)}+\omega_{(1234)}\right)}\bigg),
\eeqar
where
$F_{41}\!=\!(\omega_{(34)} + \omega_{(234)}) \left[ (\omega_{(4)}+\omega_{(34)})(\omega_{(234)}-\omega_{(1234)})+\omega_{(234)}^2-3\omega_{(234)}\omega_{(1234)} \right] - \omega_{(4)}\omega_{(234)}\omega_{(1234)}$.
\beqar
\biggl( \!\frac{dN_g^{(4)}}{dx}\! \biggl)_{2}\!&=&\!\frac{C_R}{\pi x} \int \frac{d^2 {\bf k}}{\pi}  \iiint \frac{d^2{\bf q}_{1}}{\pi} \frac{d^2{\bf q}_{2}}{\pi} \frac{d^2{\bf q}_{4}}{\pi}
\nonumber \\ &&
\alpha_{s}(Q^{2}_k) \frac{1}{\lambda^4_{dyn}}
\frac{\mu_{E}^2-\mu_{M}^2}{(\qT^2 + \mu_{E}^2)(\qT^2 + \mu_{M}^2)}
\frac{\mu_{E}^2-\mu_{M}^2}{(\qTd^2 + \mu_{E}^2)(\qTd^2 + \mu_{M}^2)}
\frac{\mu_{E}^2-\mu_{M}^2}{(\qTc^2 + \mu_{E}^2)(\qTc^2 + \mu_{M}^2)} \nonumber\\ &&
\frac{\chi^2 (\qTc\cdot(\qT+\qTd+\qTc-\kT))+(\qTc\cdot\kT)(\kT-\qTc)^2 +
(\kT\cdot(\qT+\qTd))(\qTc\cdot(\qTc-2\kT))+\kT^2(\qTc\cdot(\qT+\qTd)) }{(\chi^2+\kT^2)(\chi^2+(\kT-\qTc)^2)(\chi^2+(\kT-\qT-\qTd-\qTc)^2)}
\nonumber \\ &&
\bigg(
\frac{2 \left(\omega_{(24)} \left(2 \omega_{(4)}^2+3 \omega
   _{(24)} \omega_{(4)}+\omega_{(24)}^2\right)-\left(2 \omega_{(4)}^2+6 \omega_{(24)} \omega_{(4)}+3 \omega_{(24)}^2\right) \omega
   _{(124)}\right) \cos \left(L \left(2 \omega_{(4)}+\omega_{(24)}\right)\right)}{\omega_{(24)}^2 \left(\omega_{(4)}+\omega
   _{(24)}\right){}^2 \left(2 \omega_{(4)}+\omega_{(24)}\right){}^2 \omega_{(124)}}
\nonumber \\ &&
-\frac{2 \cos \left(L \left(2 \omega_{(4)}+\omega_{(24)}+\omega
   _{(124)}\right)\right)}{\omega_{(124)} \left(\omega_{(24)}+\omega_{(124)}\right) \left(\omega_{(4)}+\omega_{(24)}+\omega
   _{(124)}\right) \left(2 \omega_{(4)}+\omega_{(24)}+\omega_{(124)}\right)}
+\frac{\omega_{(124)} \cos \left(2 L \omega
   _{(4)}\right)}{\omega_{(4)}^2 \omega_{(24)}^2 \left(\omega_{(24)}+\omega_{(124)}\right)}
\nonumber \\ &&
-\frac{2 L \sin \left(L \left(2 \omega_{(4)}+\omega_{(24)}\right)\right)}{\omega_{(24)} \left(\omega_{(4)}+\omega
   _{(24)}\right) \left(2 \omega_{(4)}+\omega_{(24)}\right)}
-\frac{2 \omega_{(124)} \cos \left(L \omega
   _{(4)}\right)}{\omega_{(4)}^2 \left(\omega_{(4)}+\omega_{(24)}\right){}^2 \left(\omega_{(4)}+\omega_{(24)}+\omega_{(124)}\right)}
\nonumber \\ &&
+\frac{   \omega_{(124)}}{\omega_{(4)}^2 \left(2 \omega_{(4)}+\omega_{(24)}\right){}^2 \left(2 \omega_{(4)}+\omega_{(24)}+\omega
   _{(124)}\right)}\bigg),
\eeqar
\beqar
\biggl( \frac{dN_g^{(4)}}{dx} \biggl)_{3} \!&=&\!  \frac{C_R}{2\pi x} \int \frac{d^2 {\bf k}}{\pi}  \iiint \frac{d^2{\bf q}_{1}}{\pi} \frac{d^2{\bf q}_{3}}{\pi} \frac{d^2{\bf q}_{4}}{\pi}
\nonumber \\ &&
 \alpha_{s}(Q^{2}_k) \frac{1}{\lambda^4_{dyn}}
\frac{\mu_{E}^2-\mu_{M}^2}{(\qT^2 + \mu_{E}^2)(\qT^2 + \mu_{M}^2)}
\frac{\mu_{E}^2-\mu_{M}^2}{(\qTt^2 + \mu_{E}^2)(\qTt^2 + \mu_{M}^2)}
\frac{\mu_{E}^2-\mu_{M}^2}{(\qTc^2 + \mu_{E}^2)(\qTc^2 + \mu_{M}^2)}\nonumber \\ &&
\frac{\chi^2(\qTc\cdot(\qT+\qTt+\qTc-\kT))+(\qTc\cdot\kT)(\kT-\qTc)^2
+(\kT\cdot(\qT+\qTt))(\qTc\cdot(\qTc-2\kT))+\kT^2(\qTc\cdot(\qT+\qTt))}{(\chi^2+\kT^2)(\chi^2+(\kT-\qTc)^2)(\chi^2+(\kT-\qT-\qTt-\qTc)^2)}
\nonumber \\ &&
\biggl(-\frac{2 L \sin \left(L \left(\omega_{(4)}+2 \omega_{(34)}\right)\right)}{\omega_{(34)}^2 \left(\omega_{(4)}+2
   \omega_{(34)}\right)}
-\frac{4 \cos \left(L \left(\omega_{(4)}+2 \omega_{(34)}+\omega_{(134)}\right)\right)}{\omega_{(134)}
   \left(\omega_{(34)}+\omega_{(134)}\right) \left(2 \omega_{(34)}+\omega_{(134)}\right) \left(\omega_{(4)}+2 \omega_{(34)}+\omega
   _{(134)}\right)}
\nonumber \\ &&
+\frac{\left(2 \omega_{(34)} \left(\omega_{(4)}+2 \omega_{(34)}\right)-\left(3 \omega_{(4)}+8 \omega_{(34)}\right)
   \omega_{(134)}\right) \cos \left(L \left(\omega_{(4)}+2 \omega_{(34)}\right)\right)}{\omega_{(34)}^3 \left(\omega_{(4)}+2 \omega
   _{(34)}\right){}^2 \omega_{(134)}}
-\frac{\omega_{(134)} \cos \left(L
   \omega_{(4)}\right)}{\omega_{(4)} \omega_{(34)}^3(2 \omega_{(34)}+\omega_{(134)})}
\nonumber \\ &&
+\frac{4 \omega_{(134)}}{\left(\omega_{(4)}+\omega_{(34)}\right)}
\left(\frac{\cos \left(L \left(\omega_{(4)}+\omega_{(34)}\right)\right)}{\omega
   _{(34)}^3 \left(\omega_{(34)}+\omega_{(134)}\right)}
+\frac{1}{\omega_{(4)}
   \left(\omega_{(4)}+2 \omega_{(34)}\right){}^2 \left(\omega_{(4)}+2 \omega_{(34)}+\omega
   _{(134)}\right)}\right)\biggl),
\eeqar

\beqar
\biggl( \frac{dN_g^{(4)}}{dx} \biggl)_{4} \!&=&\!  \frac{C_R}{3\pi x} \int \frac{d^2 {\bf k}}{\pi} \iint \frac{d^2{\bf q}_{1}}{\pi} \frac{d^2{\bf q}_{4}}{\pi}
 \alpha_{s}(Q^{2}_k) \frac{1}{\lambda^4_{dyn}}
\frac{\mu_{E}^2-\mu_{M}^2}{(\qT^2 + \mu_{E}^2)(\qT^2 + \mu_{M}^2)}
\frac{\mu_{E}^2-\mu_{M}^2}{(\qTc^2 + \mu_{E}^2)(\qTc^2 + \mu_{M}^2)}
\nonumber \\ &&
\frac{\chi^2(\qTc\cdot(\qT+\qTc-\kT))+(\qTc\cdot\kT)(\kT-\qTc)^2+(\kT\cdot\qT)(\qTc\cdot(\qTc-2\kT))+\kT^2(\qTc\cdot\qT) }{(\chi^2+\kT^2)(\chi^2+(\kT-\qTc)^2)(\chi^2+(\kT-\qT-\qTc)^2)}
\nonumber \\ &&
\frac{1}{\omega_{(4)}^3} \bigg(-\frac{6 \omega_{(4)}^3 \cos \left(L \left(3 \omega_{(4)}+\omega
   _{(14)}\right)\right)}{\omega_{(14)} \left(\omega_{(4)}+\omega_{(14)}\right) \left(2 \omega_{(4)}+\omega_{(14)}\right) \left(3 \omega
   _{(4)}+\omega_{(14)}\right)}
+ \left(\frac{1}{\omega_{(14)}} - \frac{11}{6\omega_{(4)}} \right) \cos (3 L\omega_{(4)} )
\nonumber \\ &&
 -L \sin \left(3 L\omega_{(4)}\right)-\frac{3 \omega_{(14)} \cos
   \left(L\omega_{(4)}\right)}{4 \omega_{(4)}^2+2 \omega_{(14)} \omega_{(4)}}+\frac{3 \omega_{(14)} \cos \left(2 L\omega_{(4)}\right)}{\left(\omega_{(4)}+\omega_{(14)}\right)
   \omega_{(4)}}+\frac{\omega_{(14)}}{9 \omega_{(4)}^2+3 \omega_{(14)} \omega_{(4)}} \bigg),
\eeqar
\beqar
\biggl( \!\frac{dN_g^{(4)}}{dx} \biggl)_{5} \!&=&\!  \frac{C_R}{\pi x} \int \frac{d^2 {\bf k}}{\pi}  \iiint \frac{d^2{\bf q}_{2}}{\pi} \frac{d^2{\bf q}_{3}}{\pi} \frac{d^2{\bf q}_{4}}{\pi}
\nonumber \\ &&
 \alpha_{s}(Q^{2}_k) \frac{1}{\lambda^4_{dyn}}
\frac{\mu_{E}^2-\mu_{M}^2}{(\qTd^2 + \mu_{E}^2)(\qTd^2 + \mu_{M}^2)}
\frac{\mu_{E}^2-\mu_{M}^2}{(\qTt^2 + \mu_{E}^2)(\qTt^2 + \mu_{M}^2)}
\frac{\mu_{E}^2-\mu_{M}^2}{(\qTc^2 + \mu_{E}^2)(\qTc^2 + \mu_{M}^2)}
\nonumber \\ &&
\frac{\chi^2(\qTc\cdot(\qTd+\qTt+\qTc-\kT))+(\qTc\cdot\kT)(\kT-\qTc)^2
+(\kT\cdot(\qTd+\qTt))(\qTc\cdot(\qTc-2\kT))+\kT^2(\qTc\cdot(\qTd+\qTt))}{(\chi^2+\kT^2)(\chi^2+(\kT-\qTc)^2)(\chi^2+(\kT-\qTd-\qTt-\qTc)^2)}
\nonumber \\ &&
\frac{1}{\omega _{(234)}^2} \bigg(
\frac{2 \omega_{(234)}^3}{\omega_{(4)} \left(\omega
   _{(4)}+\omega_{(34)}\right) \left(\omega_{(4)}+\omega_{(34)}+\omega_{(234)}\right){}^2 \left(\omega_{(4)}+\omega_{(34)}+2 \omega_{(234)}\right)}
+\frac{\cos \left(L \left(\omega_{(4)}+\omega
   _{(34)}\right)\right)}{\omega_{(34)} \left(\omega_{(4)}+\omega_{(34)}\right)}
\nonumber \\ &&
-\frac{2 \omega_{(234)}^3 \cos \left(L \omega_{(4)}\right)}{\omega_{(4)} \omega_{(34)} \left(\omega_{(34)}+\omega
   _{(234)}\right){}^2 \left(\omega_{(34)}+2 \omega_{(234)}\right)}
- \frac{2L\omega_{(234)} \sin(L(\omega_{(4)}+\omega_{(34)}+\omega_{(234)}))}{\left(\omega_{(34)}+\omega
   _{(234)}\right) \left(\omega_{(4)}+\omega_{(34)}+\omega_{(234)}\right)}
\nonumber \\ &&
\!-\frac{\cos \left(L \left(\omega_{(4)}\!+\!\omega_{(34)}+2
   \omega_{(234)}\right)\right)}{\left(\omega_{(34)}\!+\!2 \omega_{(234)}\right) \left(\omega_{(4)}\!+\!\omega_{(34)}+2 \omega
   _{(234)}\right)}
\!-\!\frac{2\omega_{(234)} (\omega_{(4)}\!+\!2\omega_{(34)}\!+\!2\omega_{(234)}) \cos(L(\omega_{(4)}\!+\!\omega_{(34)}\!+\!\omega_{(234)}))}{\left(\omega_{(34)}\!+\!\omega_{(234)}\right)^2 \left(\omega_{(4)}\!+\!\omega_{(34)}\!+\!\omega_{(234)}\right)^2}
\! \bigg),\nonumber \\ &&
\eeqar
\beqar
\biggl( \frac{dN_g^{(4)}}{dx} \biggl)_{6} \!&=&\! \frac{C_R}{\pi x} \int \frac{d^2 {\bf k}}{\pi}  \iint \frac{d^2{\bf q}_{2}}{\pi} \frac{d^2{\bf q}_{4}}{\pi}
 \alpha_{s}(Q^{2}_k) \frac{1}{\lambda^4_{dyn}}
\frac{\mu_{E}^2-\mu_{M}^2}{(\qTd^2 + \mu_{E}^2)(\qTd^2 + \mu_{M}^2)}
\frac{\mu_{E}^2-\mu_{M}^2}{(\qTc^2 + \mu_{E}^2)(\qTc^2 + \mu_{M}^2)}
\nonumber \\ &&
\frac{\chi^2(\qTc\cdot(\qTd+\qTc-\kT))+(\qTc\cdot\kT)(\kT-\qTc)^2+(\kT\cdot\qTd)(\qTc\cdot(\qTc-2\kT))+\kT^2(\qTc\cdot\qTd)  }{(\chi^2+\kT^2)(\chi^2+(\kT-\qTc)^2)(\chi^2+(\kT-\qTd-\qTc)^2)}
\nonumber \\ &&
\bigg(\frac{\omega_{(24)}}{8\omega_{(4)}^2 \left(\omega_{(4)}+\frac{\omega_{(24)}}{2}\right){}^2 \left(\omega_{(4)}+\omega
   _{(24)}\right)}
-\frac{L \sin \left(L \left(2
   \omega_{(4)}+\omega_{(24)}\right)\right)+\frac{\left(\frac{3 \omega_{(4)}}{2}+\omega_{(24)}\right) \cos \left(2 L \left(\omega
   _{(4)}+\frac{\omega_{(24)}}{2}\right)\right)}{\left(\omega_{(4)}+\frac{\omega_{(24)}}{2}\right) \left(\omega_{(4)}+\omega
   _{(24)}\right)}}{\left(\omega_{(4)}+\frac{\omega_{(24)}}{2}\right) \left(\omega_{(4)}+\omega_{(24)}\right) \omega
   _{(24)}}
\nonumber \\ &&
+\frac{\left(\frac{\cos \left(2 L \omega_{(4)}\right)}{2 \omega
   _{(4)}}-\frac{\omega_{(4)} \cos \left(2 L \left(\omega_{(4)}+\omega_{(24)}\right)\right)}{4 \left(\frac{\omega_{(4)}}{2}+\omega
   _{(24)}\right) \left(\omega_{(4)}+\omega_{(24)}\right)}\right)}{\omega_{(4)} \omega_{(24)}^2}
-\frac{\omega_{(24)} \cos \left(L \omega_{(4)}\right)}{\omega_{(4)}^2 \left(\frac{\omega_{(4)}}{2}+\omega
   _{(24)}\right) \left(\omega_{(4)}+\omega_{(24)}\right){}^2}
\bigg),
\eeqar

\beqar
\biggl(\! \frac{dN_g^{(4)}}{dx} \biggl)_{7} \!&=&\!  \frac{C_R}{2\pi x}  \int \frac{d^2 {\bf k}}{\pi}  \iint \frac{d^2{\bf q}_{3}}{\pi} \frac{d^2{\bf q}_{4}}{\pi}
 \alpha_{s}(Q^{2}_k) \frac{1}{\lambda^4_{dyn}}
\frac{\mu_{E}^2-\mu_{M}^2}{(\qTt^2 + \mu_{E}^2)(\qTt^2 + \mu_{M}^2)}
\frac{\mu_{E}^2-\mu_{M}^2}{(\qTc^2 + \mu_{E}^2)(\qTc^2 + \mu_{M}^2)}  \nonumber \\ &&
\frac{\chi^2(\qTc\cdot(\qTt+\qTc-\kT))+(\qTc\cdot\kT)(\kT-\qTc)^2 + (\kT\cdot\qTt)(\qTc\cdot(\qTc-2\kT))+\kT^2(\qTc\cdot\qTt)}{(\chi^2+\kT^2)(\chi^2+(\kT-\qTc)^2)(\chi^2+(\kT-\qTt-\qTc)^2)}
\nonumber \\ &&
\frac{1}{\omega_{(34)}^2 \!\left(\frac{\omega_{(4)}}{2}\!+\!\omega_{(34)}\right)}
\bigg(\frac{2 \omega_{(34)}^3}{\omega_{(4)}
   \left(\omega_{(4)}\!+\!\omega_{(34)}\right) \left(\omega_{(4)}\!+\!2 \omega_{(34)}\right) \left(\omega_{(4)}\!+\!3 \omega_{(34)}\right)}
\!-\!\frac{\left(\frac{\omega_{(4)}}{2}\!+\!2 \omega_{(34)}\right) \cos \left(L \left(\omega_{(4)}\!+\!2
   \omega_{(34)}\right)\right)}{\left(\omega_{(4)}\!+\!2 \omega_{(34)}\right) \omega_{(34)}}
 \nonumber \\ &&
 -L \sin \left(L \left(\omega_{(4)}\!+\!2 \omega_{(34)}\right)\right)\!-\!\frac{\left(\frac{\omega_{(4)}}{2}\!+\!\omega_{(34)}\right)
   \left(\frac{\cos \left(L \omega_{(4)}\right)}{\omega_{(4)}}\!-\!\frac{6 \cos \left(L \left(\omega_{(4)}\!+\!\omega
   _{(34)}\right)\right)}{\omega_{(4)}\!+\!\omega_{(34)}}\!+\!\frac{2 \cos \left(L \left(\omega_{(4)}\!+\!3 \omega_{(34)}\right)\right)}{\omega_{(4)}\!+\!3
   \omega_{(34)}}\right)}{3 \omega_{(34)}}\bigg),
\eeqar
\beqar
\biggl( \frac{dN_g^{(4)}}{dx} \biggl)_{8} \!&=&\!  \frac{C_R}{3\pi x} \int \frac{d^2 {\bf k}}{\pi}  \int \frac{d^2{\bf q}_{4}}{\pi}
 \alpha_{s}(Q^{2}_k) \frac{1}{\lambda^4_{dyn}}
\frac{\mu_{E}^2-\mu_{M}^2}{(\qTc^2 + \mu_{E}^2)(\qTc^2 + \mu_{M}^2)}
\frac{\chi^2\qTc\cdot(\qTc-\kT) + (\qTc\cdot\kT)(\kT-\qTc)^2}{(\chi^2+\kT^2)(\chi^2+(\kT-\qTc)^2)^2}
\nonumber \\ &&
\frac{1}{\omega_{(4)}^3} \!\left(\! \frac{1}{12 \omega_{(4)}}\!-\!L \sin\! \left(3 L \omega_{(4)}\right)\!-\!\frac{\cos\! \left(L \omega_{(4)}\right)}{2 \omega_{(4)}}\!+\!\frac{3 \cos \!\left(2 L \omega
   _{(4)}\right)}{2 \omega_{(4)}}\!-\!\frac{5 \cos \!\left(3 L \omega_{(4)}\right)}{6 \omega_{(4)}}\!-\!\frac{\cos \!\left(4 L \omega_{(4)}\right)}{4
   \omega_{(4)}} \!\right).
\eeqar

\section{$dN_g/dx$ results for $L=3$ and $L=1$}

In this section, we show $dN_g/dx$ as a function of $x$ for medium lengths $L=3fm$ (Fig.~(\ref{fig:dNdxL3})) and $L=1fm$ (Fig.~(\ref{fig:dNdxL1})).

\begin{figure*}
\centering
\epsfig{file=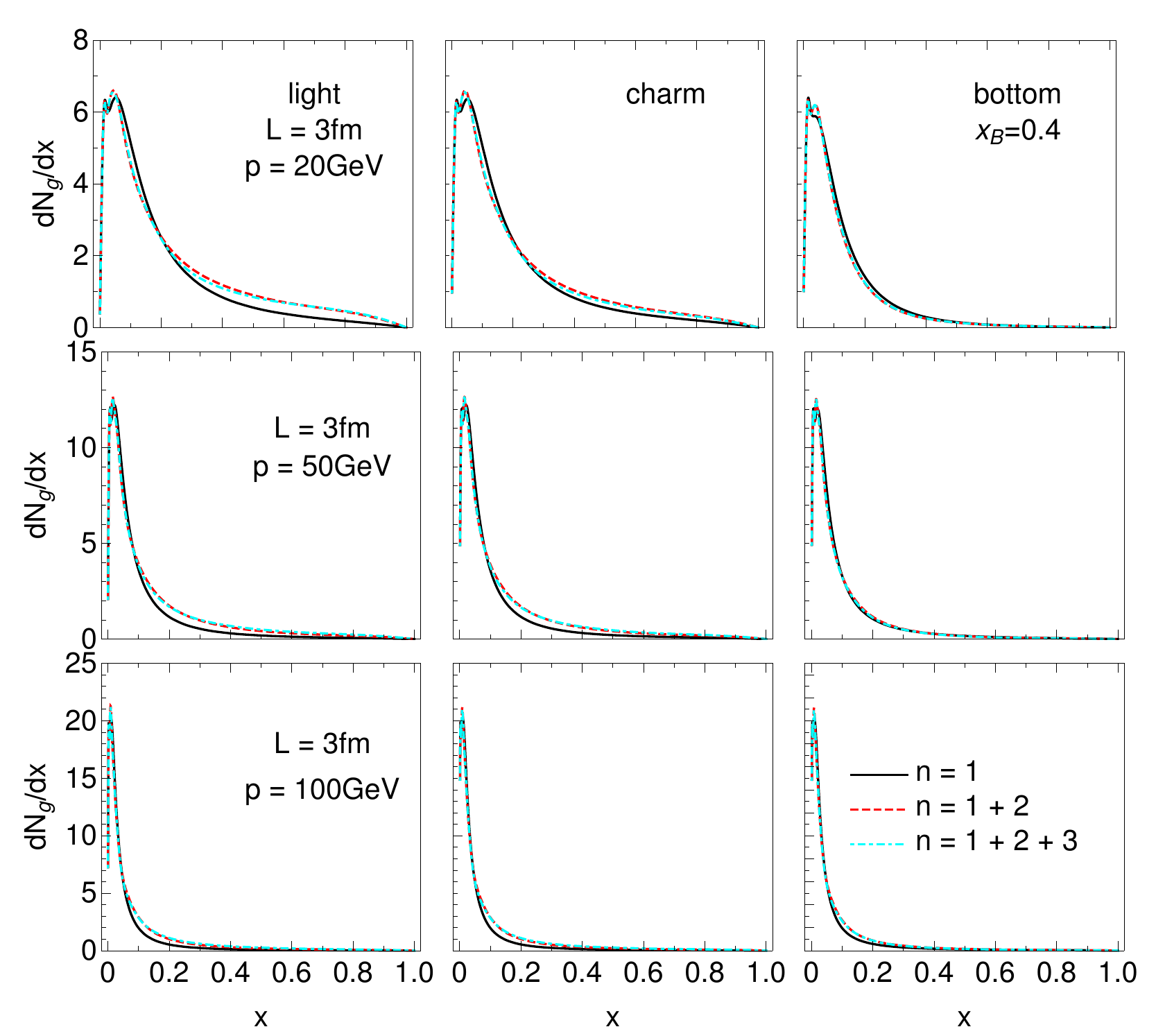,scale=0.3}\epsfig{file=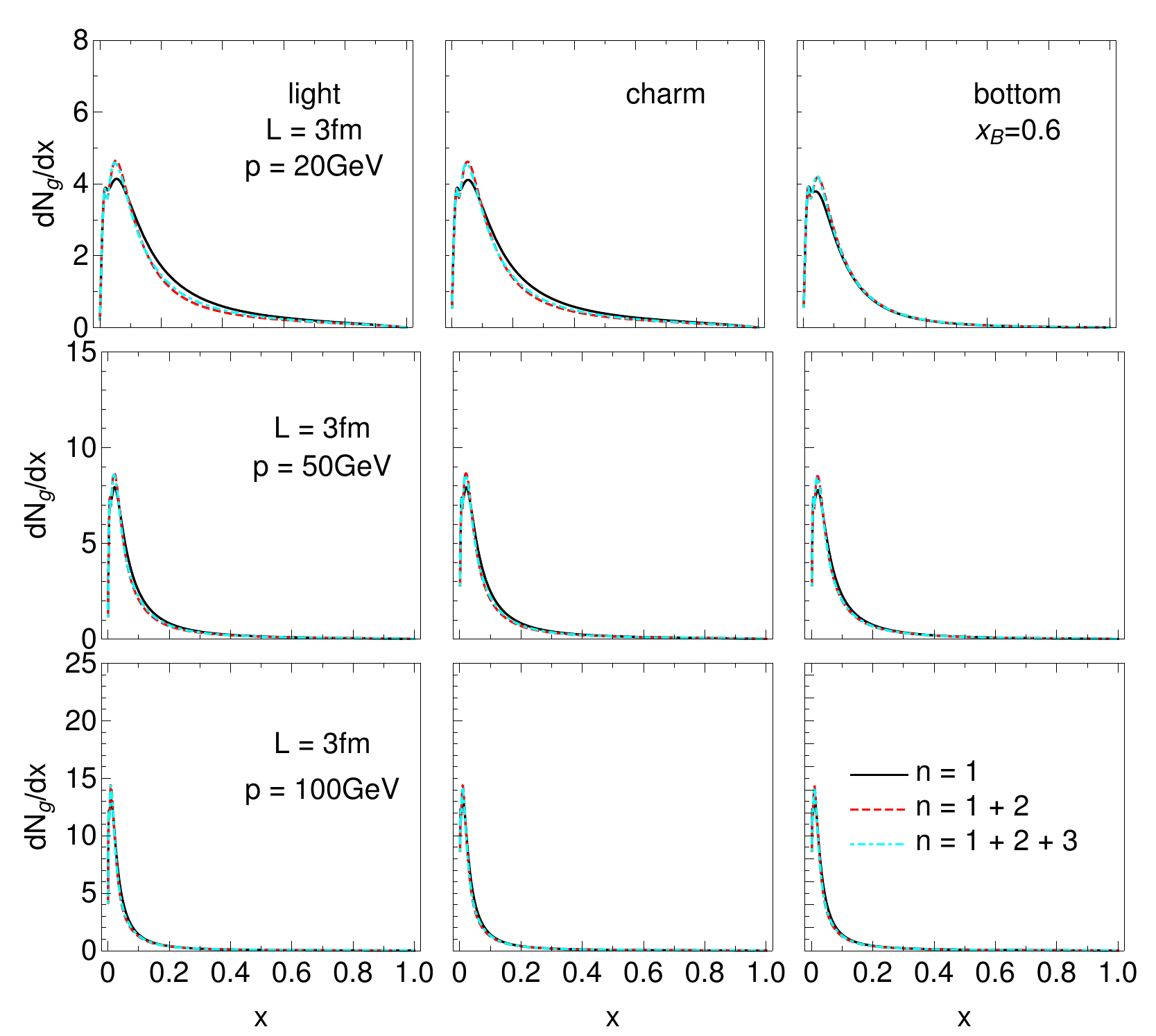,scale=0.3}
\vspace*{-0.2cm}
\caption{Gluon radiation spectrum $dN_g/dx$ as a function of $x$, for the medium length of $L=3fm$ and various jet momenta. The panel on the left (right) side shows the result for $\mu_M / \mu_E = 0.4$ $(0.6)$. The figure caption is the same as for Fig. 1.}
\label{fig:dNdxL3}
\end{figure*}

\begin{figure*}
\centering
\epsfig{file=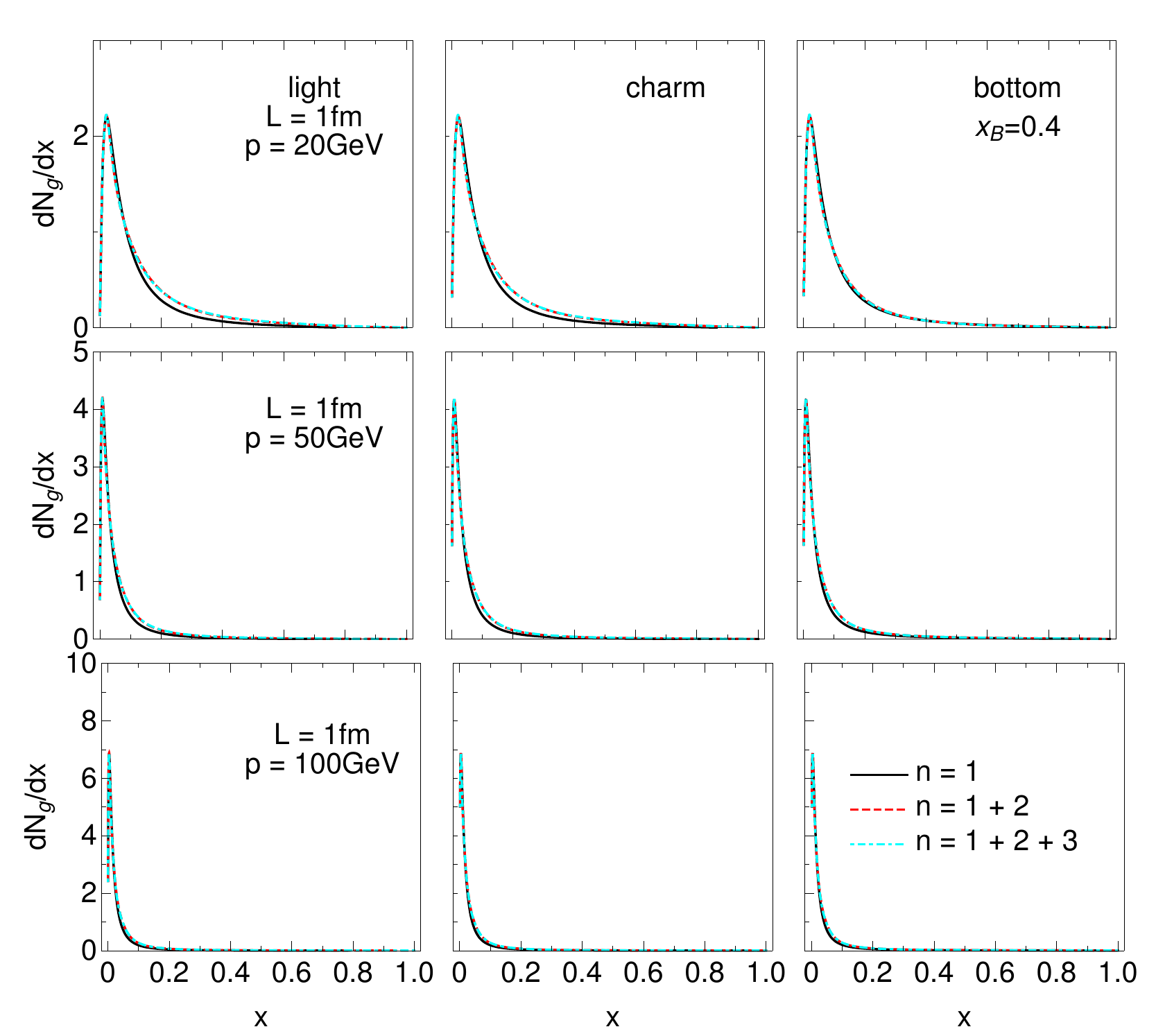,scale=0.3}\epsfig{file=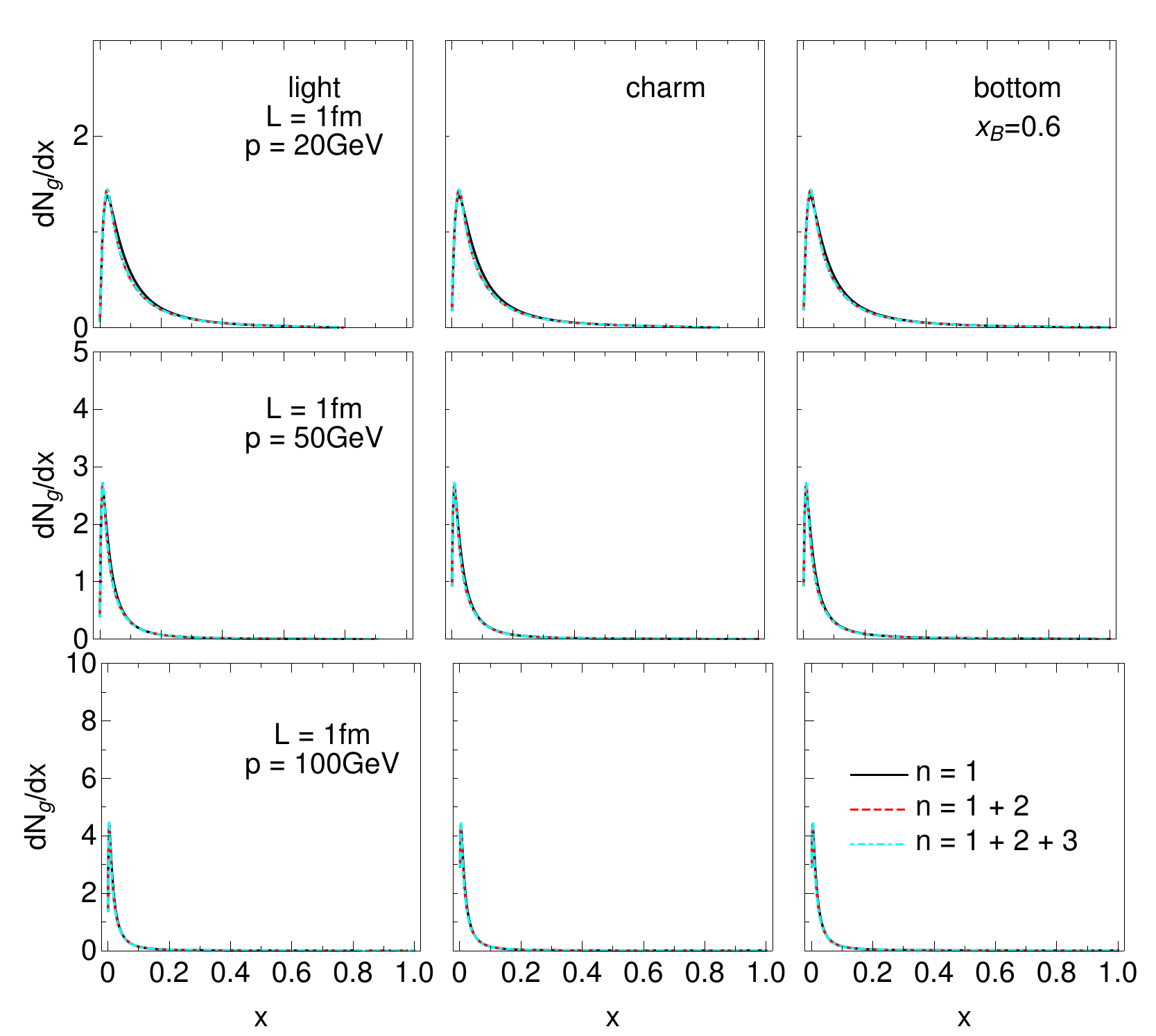,scale=0.3}
\vspace*{-0.2cm}
\caption{Gluon radiation spectrum $dN_g/dx$ as a function of $x$, for the medium length of $L=1fm$ and various jet momenta. The panel on the left (right) side shows the result for $\mu_M / \mu_E = 0.4$ $(0.6)$. The figure caption is the same as for Fig. 1.}
\label{fig:dNdxL1}
\end{figure*}

\section{$v_2$ results up to the 3$^{rd}$ order in opacity}

We here show the results for $v_2$ up to the 3$^{rd}$ order in opacity (Fig.~\ref{fig:v2}). Note that here the lower (upper) boundary of each band corresponds to the $\mu_M / \mu_E = 0.6$ ($\mu_M / \mu_E = 0.4$) case (opposite with respect to $R_{AA}$ results). We observe the same behavior as for $R_{AA}$.

\begin{figure*}
\centering
\epsfig{file=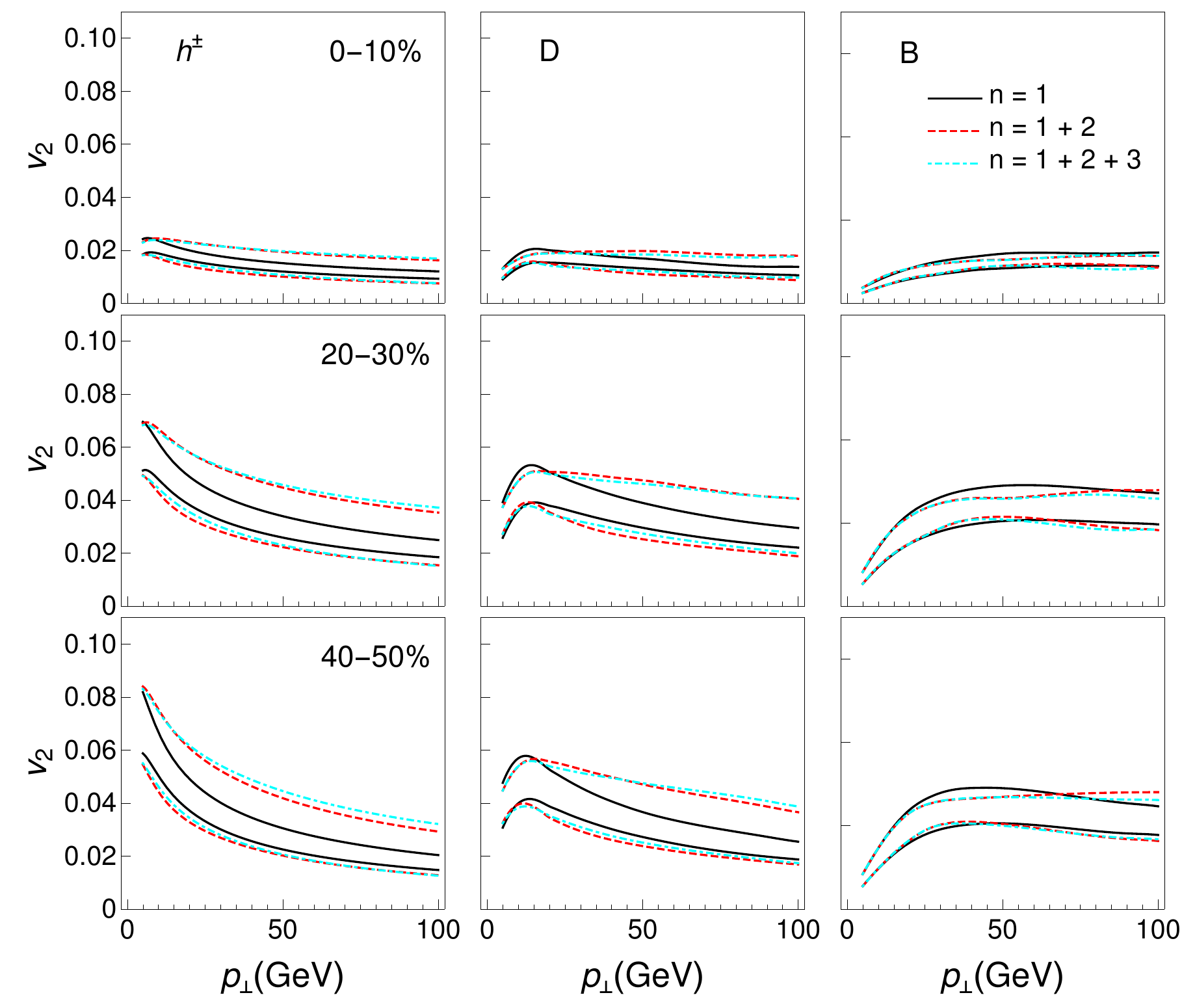,scale=0.35}
\vspace*{-0.2cm}
\caption{$v_2$ results obtained within DREENA-C -- the effects of different orders in opacity. Different columns correspond to charged hadrons, D, and B mesons, while different rows show different centrality classes. Only radiative energy loss is taken into account. Solid black curves show the $1^{\mathrm{st}}$ order in opacity results, red dashed curves show the results up to the $2^{\mathrm{nd}}$ order, while cyan dot-dashed curves up to the $3^{\mathrm{rd}}$ order in opacity. The lower (upper) boundary of each band corresponds to the $\mu_M / \mu_E = 0.6$ ($\mu_M / \mu_E = 0.4$) case.}
\label{fig:v2}
\end{figure*}

\endgroup

\end{document}